\documentclass[prd,amsmath,amssymb,showpacs,floatfix,
superscriptaddress,nofootinbib,preprintnumbers]{revtex4}
\usepackage[latin1]{inputenc}
\usepackage{amsmath}
\usepackage{amsfonts}
\usepackage{amssymb}
\usepackage{graphicx}
\usepackage{appendix}
\usepackage{dcolumn}
\usepackage{caption}
\usepackage{subfigure}
\usepackage{float}
\usepackage{bm}
\usepackage{latexsym}
\usepackage{epsfig}

\begin{document}
\title{Statistics of Bipolar Representation of CMB maps.}
\author{Nidhi Joshi\footnote{nidhijoshi@ctp-jamia.res.in}}
\affiliation{Centre for Theoretical Physics, Jamia Millia Islamia, New Delhi 110025, India}
\affiliation{IUCAA, Post Bag 4, Ganeshkhind, Pune-411007, India}
\author{Aditya Rotti\footnote{aditya@iucaa.ernet.in}}
\author{Tarun Souradeep\footnote{tarun@iucaa.ernet.in}}
\affiliation{IUCAA, Post Bag 4, Ganeshkhind, Pune-411007, India}
\date{\today}

\begin{abstract}
Gaussianity of temperature fluctuations in the Cosmic Microwave Background(CMB) implies that the statistical properties of the temperature field can be completely
characterized by its two point correlation function. The two point correlation function can be expanded in full generality in the bipolar spherical harmonic(BipoSH) basis. Looking for significant deviations from zero for Bipolar Spherical Harmonic(BipoSH) Coefficients derived from observed CMB maps forms the 
basis of the strategy used to detect isotropy violation. In order to quantify "significant deviation" we need to understand the distributions 
of these coefficients. We analytically evaluate the moments and the distribution of the coefficients of expansion($A^{LM}_{l_1 l_2}$), using characteristic function approach.
We show that for BipoSH coefficients with $M=0$ an analytical form for the moments up to any arbitrary order can be derived. For the remaining BipoSH coefficients with $M\neq0$, the moments derived using the characteristic function approach need to be supplemented with a correction term. The correction term is found to be important particularly at low multipoles. We provide a general prescription for calculating these corrections, however we restrict the explicit calculations only up to kurtosis. We confirm our results with measurements of BipoSH coefficients on numerically simulated statistically isotropic CMB maps.
\end{abstract}

\maketitle

\newpage		
\section{Introduction}
\indent Cosmological model building has been usually pursued under the assumption that the universe is homogeneous and isotropic.  
Statistical isotropy of CMB implies statistical expectation values of the temperature fluctuations are preserved under rotations of the sky. 
The CMB data is one of the cleanest observations and it is only reasonable to search for weak violations of statistical isotropy in the CMB maps.
The source of this deviation remains mysterious so far. However, with present as well as future maps providing more and more
information it is important to study deviations from standard statistics. 

Random temperature fluctuations in the CMB are believed to be Gaussian, these fluctuations can be completely characterized by specifying the two point correlation function.\\
\indent The two point correlation function is most generally expanded in the bipolar spherical harmonic(BipoSH) basis to test the violations of isotropy in the CMB
temperature and polarization maps. This formalism was developed by Hajian and Souradeep \cite{AH-TS-03,AH-TS-04,AH-TS-NC,AH-TS-05,SB-AH-TS,AH-TS-06} and
is such that for an isotropic sky all BipoSH coefficients, $A^{LM}_{l_1 l_2}$ except $A^{00}_{ll}$ vanish on an average. 
These expansion coefficients have been used to parametrize several kind of statistical isotropy violations \cite{MA-TS,LA-SM-MB,AP-MK,NJ-SJ-TS-AH} and was adopted by the
WMAP team \cite{CB-RH-GH} to search for violations of isotropy in the WMAP data.
Although, these coefficients were primarily 
introduced to study statistical isotropy violation, they have found various other applications \cite{MK,VS-KM-AC,MK-TS}.\\
\indent Specifying all the moments of a distribution completely characterize the distribution. In this paper, we derive analytical 
expressions for the moments of the distribution of the BipoSH coefficients
using the characteristic function approach. BipoSH coefficients are linear combinations of elements of the harmonic space covariance matrix.
The independence of the terms in 
the linear combination for the BipoSH coefficients with $M=0$ ensures that the characteristic function encodes complete statistical information. For the remaining BipoSH coefficients with $M\neq 0$,  the characteristic function method partially works due to the 
presence of non linear correlations among terms in the linear combination. To account for these non linear correlations we supply a correction term to the moments derived using the characteristic function method. We test these analytical results against simulations. We simulate the CMB maps using the widely used HealPix \cite{hpix} package.\\
\indent This paper is organized as follows. In section \ref{bipfor} we briefly discuss the BipoSH formalism introduced by Hajian and Souradeep. 
In section \ref{bipstat} we discuss the characteristic function approach which is extensively used to derive the moments of the distribution of the BipoSH coefficients. In section \ref{statbiposh}
we present the analytical expressions derived for the various BipoSH coefficients.
The details of these calculations and a detailed discussion on the characteristic function approach can be found in the appendices. 
We conclude with a discussion of our results in section \ref{conc}.

\section{BipoSH formalism}\label{bipfor}
The isotropic CMB sky is fully characterized by specifying the four angular power spectra $C_l^{TT},C_l^{BB},C_l^{EE} \textrm{ and } C_l^{TE}$. These are the Legendre polynomial coefficients of the corresponding two point correlation function defined in the following manner,
\begin{equation}\label{eq:isot}
C^{XX}(\hat{n}_{1,}\hat{n}_{2}) =C^{XX}(\hat{n}_{1.}\hat{n}_{2})=
\sum_{l}\frac{2l+1}{4\pi}C_l^{XX}P_l(\hat{n}_{1}\cdot \hat{n}_{2})\,.
\end{equation}
In what follows we drop the 'XX' label for notational brevity.\\
If the CMB  sky is not assumed to be isotropic then two point correlation function  in general will depend on the directions $\hat{n}_{1} ~\textrm{and}~\hat{n}_{2}$. 
Hence, the bipolar spherical harmonic basis form a very natural basis in which the CMB two point correlation function can be expanded,
\begin{equation}\label{eq:BPOSH}
C(\hat{n}_{1,}\hat{n}_{2}) =
\sum_{l_{1},l_{2},L,M}A_{l_{1}l_{2}}^{\ell
M}\{Y_{l_{1}}(\hat{n}_{1})\otimes Y_{l_{2}}(\hat{n}_{2})\}_{L M},
\end{equation}
where $A_{l_{1}l_{2}}^{L M}$ are BipoSH coefficients and  $\{Y_{l_{1}}(\hat{n}_{1})\otimes
Y_{l_{2}}(\hat{n}_{2})\}_{L M}$ are bipolar spherical harmonics~\cite{varshalovich}.
BipoSH functions are irreducible tensor product of two spherical harmonics with different arguments, they form an orthonormal basis on $\textbf{S}^{2} \times \textbf{S}^{2}$ for different sets 
of $l_1, l_2, L,M$. Their transformation properties under rotations are similar to spherical harmonics and can be expressed as,
\begin{equation}
\{Y_{l_{1}}(\hat{n}_{1})\otimes Y_{l_{2}}(\hat{n}_{2})\}_{L M} =
\sum_{m_{1}m_{2}} C_{l_{1}m_{1}l_{2}m_{2}}^{LM}Y^{l_{1}}_{m_{1}}(\hat{n}_{1})\;
Y^{l_{2}}_{m_{2}}(\hat{n}_{2}),
\end{equation}
where $C_{l_{1}m_{1}l_{2}m_{2}}^{LM}$ are Clebsch-Gordon coefficients. These indices satisfy triangularity conditions $|l_1-l_2|\le L\le l_1+l_2$ and $m_1+m_2=M$.
\newline \\
The BipoSH coefficients can be shown to be linear combinations of
off-diagonal elements of the harmonic space covariance matrix~\cite{AH-TS-03},
\begin{eqnarray}
A^{LM}_{l_1 l_2}=\sum_{m_1 m_2}\langle a_{l_1 m_1}a^{*}_{l_2
m_2}\rangle(-1)^{m_2}C^{LM}_{l_1 m_1 l_2 -m_2},
\end{eqnarray}
where $a_{lm}$'s are the spherical harmonic coefficients of the CMB maps.
An unbiased estimator of BipoSH coefficients can be defined in terms of the spherical harmonic coefficients of the CMB maps,
\begin{eqnarray}\label{eq:bipolar}
 A^{LM}_{l_1 l_2}=\sum_{m_1 m_2}a_{l_1 m_1}a_{l_2
m_2}C^{LM}_{l_1 m_1 l_2 m_2}\,.
\end{eqnarray}
It can be proved that for an isotropic CMB sky the expectation value of all the BipoSH coefficients, except the isotropic angular power spectrum 
$A^{0 0}_{l l}=(-1)^{l}C_{l}\sqrt{2l+1}$  vanish~\cite{AH-TS-03,NJ-SJ-TS-AH}.
\section{Characteristic function method}\label{bipstat}
We investigate the statistical properties of the real and imaginary parts of complex coefficients obtained in the BipoSH representation of the
CMB two point correlation
function. To arrive at the moments of BipoSH coefficients, which are linear combinations of covariance matrix elements (see eq.\ref{eq:BipoSH-exp}), we adopt the characteristic 
function approach which is widely used in statistics~\cite{springer}.\\
\indent The characteristic function of any random variable completely defines its probability distribution \cite{PP}. It is defined in the following manner,
\begin{eqnarray}
 \varphi_{X}(t)=E[e^{itX}]\quad\quad\quad t\in \Re \,.
\end{eqnarray}
Consider a random variable defined in the following manner,
\begin{eqnarray}\label{eq:linear}
 Z_n=\sum^{n}_{i=1}a_i X_i ,
\end{eqnarray}
where $a_i$'s are constants and $X_{i}$'s are independent random variables which are not necessarily identically distributed. By independence we imply that, all the higher order correlations between the terms appearing in the linear combination vanish,\\
\begin{eqnarray}\label{eq:independence}
 \langle X_i^n\cdot X_j^{*m} \rangle = 0 ~~~~~~~\forall ~~ n,m ~~~~~~(i \neq j) \,.
\end{eqnarray}
The characteristic function method is particularly useful in arriving at the statistics of such random variables.
The characteristic function of $Z_n$ will just be the product of the characteristic function of the individual terms contributing to the linear sum,
\begin{eqnarray}\label{eq:charprop}
 \varphi_{Z_n}(t)=\varphi_{X_1}(a_1 t)\varphi_{X_2}(a_2 t)......\varphi_{X_n}(a_n t) \,.
\end{eqnarray}
If the terms involved in the linear combination are not independent then the characteristic function will not take up the simple form given above. 
The cumulant generating function is defined as the logarithm of the characteristic function,
\begin{eqnarray}
 g_{Z}(t)=\log[\varphi_{Z}(t)] \,.
\end{eqnarray}
The cumulants can be obtained by taking derivatives of the cumulant generating function and evaluating them at zero
\begin{eqnarray}
K_n = i^{n} g^{n}_{Z}(t)|_{t=0} \,.
\end{eqnarray}
Given the cumulants, it is straightforward to arrive at the moments of the distribution. 
The explicit relationships between cumulants and central moments till the sixth central moments are given below,
\begin{eqnarray}
\mu_1 &=& K_1, \nonumber \\
\mu_2 &=& K_2, \nonumber \\
\mu_3 &=& K_3 \nonumber \\
\mu_4 &=& K_4+3 K^{2}_2, \nonumber \\
\mu_5 &=& K_5+10 K_3 K_2,  \nonumber \\
\mu_6 &=& K_6 + 15 K_4 K_2 +10 K^{2}_3 +15 K^{3}_2 \,.
\end{eqnarray}
Each term in the expansion for moments in terms of the cumulants is of the form $K^{A}_{a}*K^{B}_{b}*K^{C}_{c}*....$, such that $aA+bB+c C+....=n$.\\
Also note that, $A,B,C...\geq 1$ and 
$2\leq a,b,c,...\leq n$, where $n$ is the moment that one is interested in.
The coefficient of any general term in the expansion of the moment in terms of the cumulant is given by,
\[
\frac{n!}{A!*{a!}^{A}*B!*{b!}^{B}*C!*{c!}^{C}....}\,.
\]
Note that in the figures that appear in the rest of the article we plot the normalized moments defined by,
\begin{eqnarray}\label{eq:moments}
\mu^{Norm}_n=\frac{\mu_n}{\sigma^n}\,.
\end{eqnarray}
\section{Statistics of Bipolar spherical harmonic coefficients}\label{statbiposh}
\indent We classify BipoSH coefficients into four different cases depending upon the form of their characteristic function.
\medskip
\newline\textbf{Case A}: $l_1 = l_2, M=0$,
\newline\textbf{Case B}: $l_1\ne l_2, M = 0$,
\newline\textbf{Case C}: $l_1 = l_2, M\ne 0$,
\newline\textbf{Case D}: $l_1\ne l_2, M\ne 0$.
\medskip 
\newline To arrive at the distribution of a given BipoSH coefficient, we begin with finding out the characteristic function of individual terms involved in the linear combination. 
Now, the characteristic function of the BipoSH coefficients can be written as the product of the characteristic functions of each of the individual terms present in summation. \\
\indent This simple scheme works really well for cases {\bf A} and {\bf B}. However it only partially works in the the cases {\bf C} and {\bf D} as there appear terms in the 
summation which are not statistically independent of each other. 
For these cases, we calculate the moments using the method of characteristic function and then present a general prescription for calculating the correction to these 
moments. All the moments are evaluated in units of standard deviation ($\sigma$).
\subsection*{Case A: Bipolar coefficient with $l_1 = l_2=l,M= 0$}
\begin{figure}[!hbtp]
	\centering
    \subfigure[Std. Dev. ($\sigma$)]{\label{fig:edge-a}  \includegraphics[height=5.8cm,width=4cm,angle=-90]{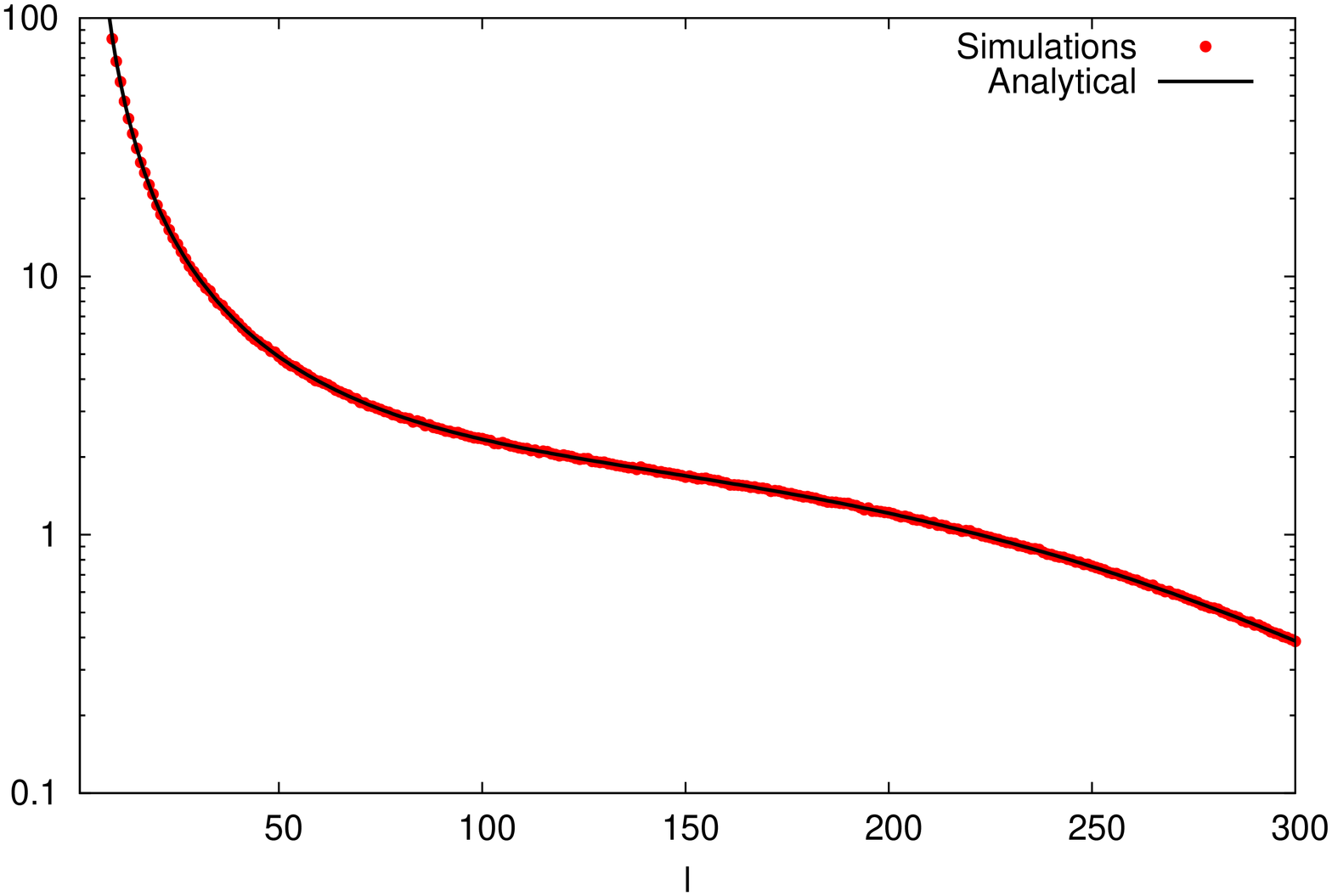}}
	\subfigure[Kurtosis] {\label{fig:edge-a} \includegraphics[height=5.8cm,width=4cm,angle=-90]{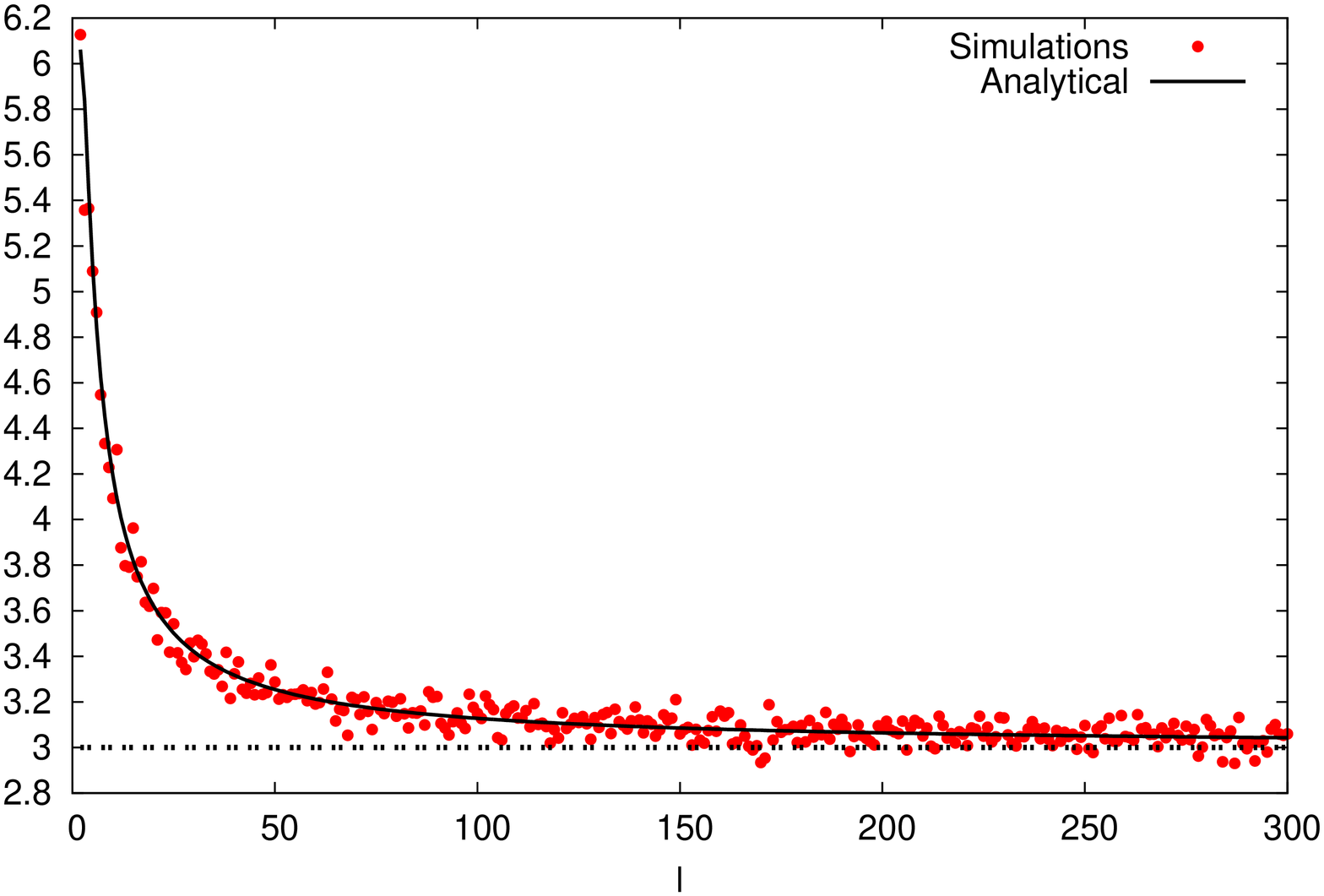}}
	\subfigure[Skewness]{\label{fig:edge-a}  \includegraphics[height=5.8cm,width=4cm,angle=-90]{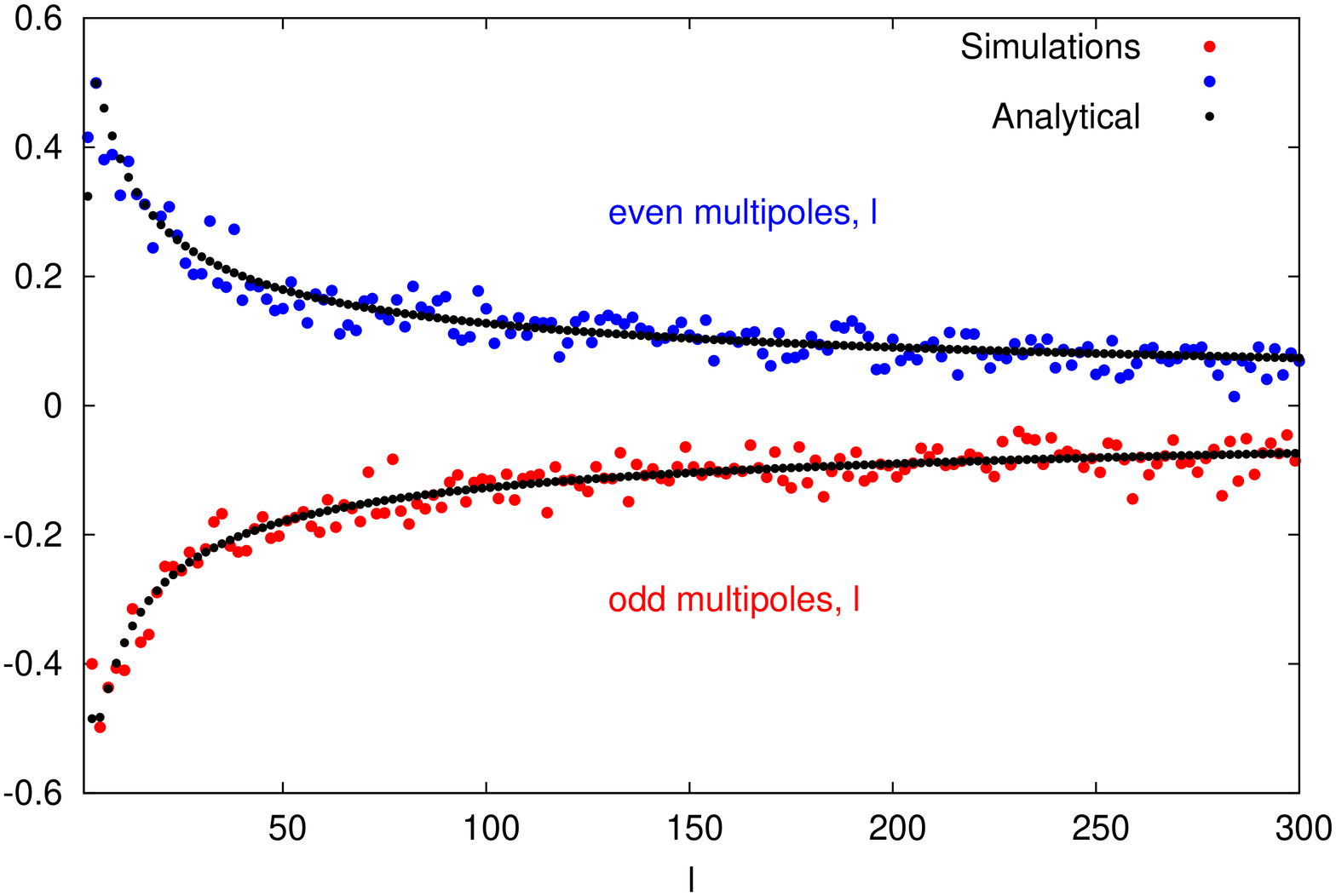}}
	\subfigure[$5^{th}$ Moment] {\label{fig:edge-a} \includegraphics[height=5.8cm,width=4cm,angle=-90]{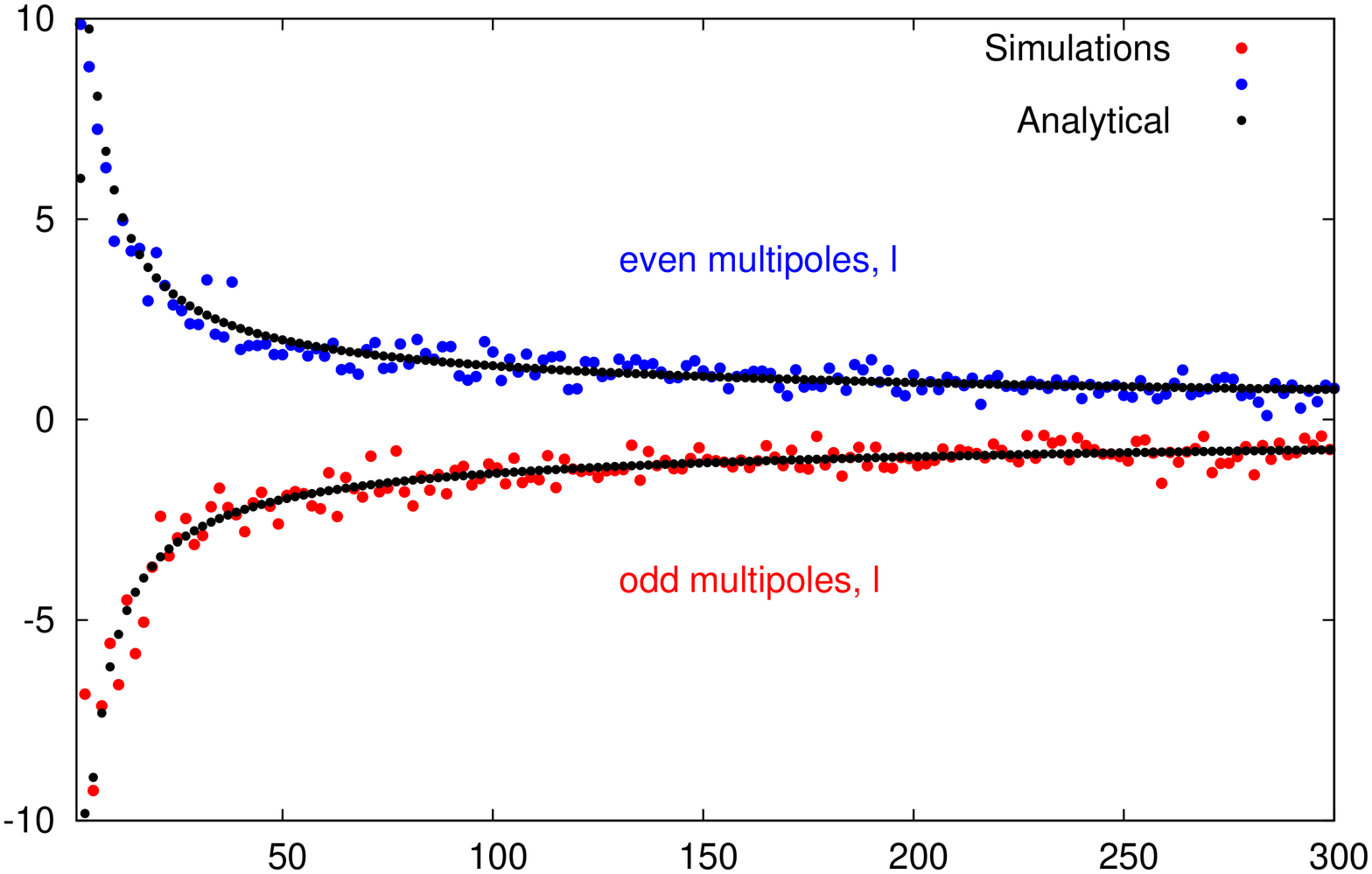}}
	\vskip -3mm
	 \caption{Standard deviation, skewness, kurtosis and $5^{th}$ moment of real part of $A^{20}_{l l}$, from 15000 simulations. WMAP7 has detected a signal of isotropy violation in these coefficients around the multipole of 200. Hence we calculate the statistics of these coefficients up to higher multipoles.}
	  \label{fig4}
\end{figure}

\begin{figure}[!ht]
	\centering
    \subfigure[Mean]{\label{fig:edge-a}  \includegraphics[height=5.8cm,width=4cm,angle=-90]{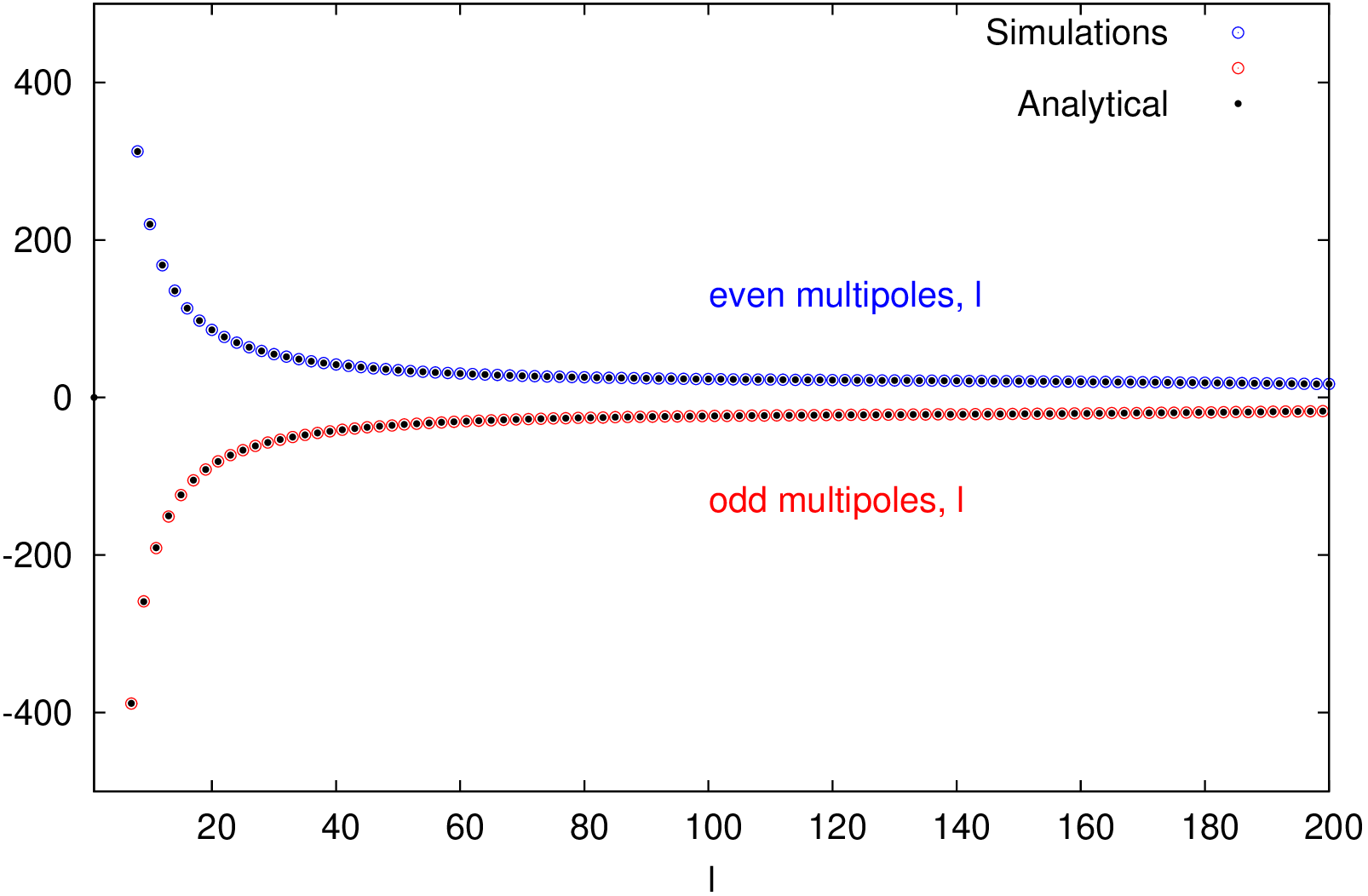}}
	\subfigure[$A^{00}_{66}$] {\label{fig:edge-a} \includegraphics[height=5.8cm,width=4cm,angle=-90]{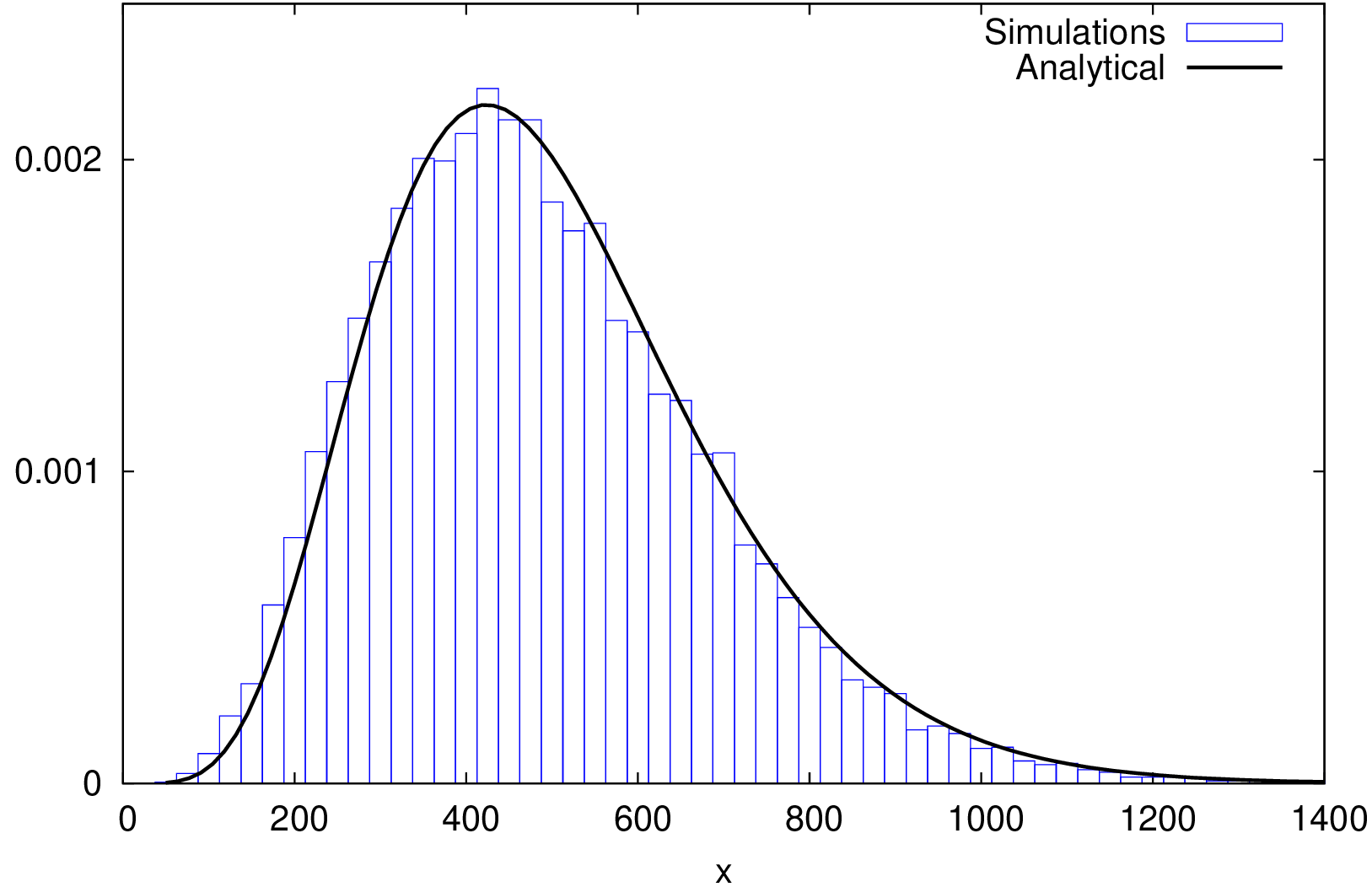}}
	\subfigure[$A^{00}_{7070}$]{\label{fig:edge-a}  \includegraphics[height=5.8cm,width=4cm,angle=-90]{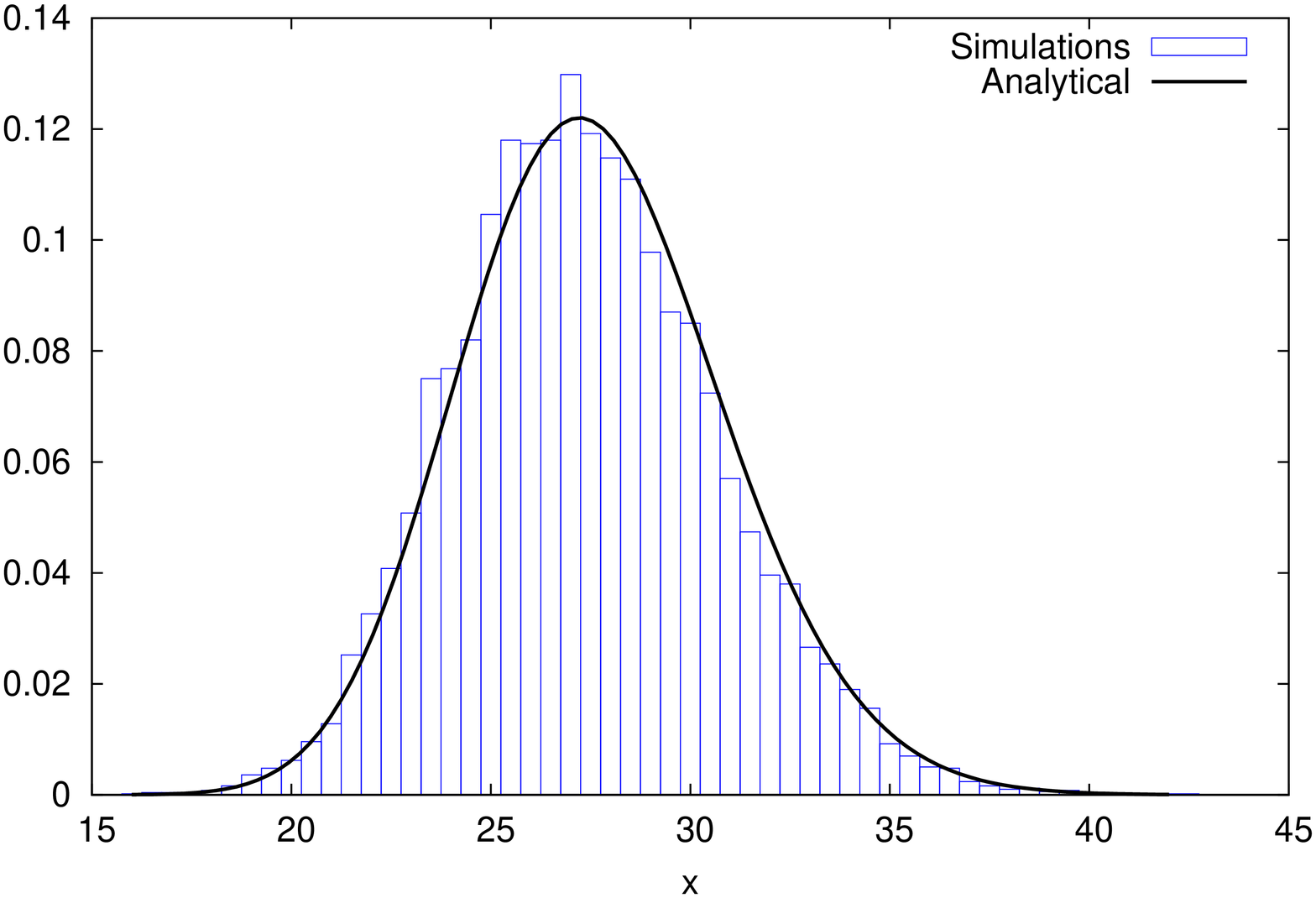}}
	\subfigure[$A^{00}_{1111}$ Moment] {\label{fig:edge-a} \includegraphics[height=5.8cm,width=4cm,angle=-90]{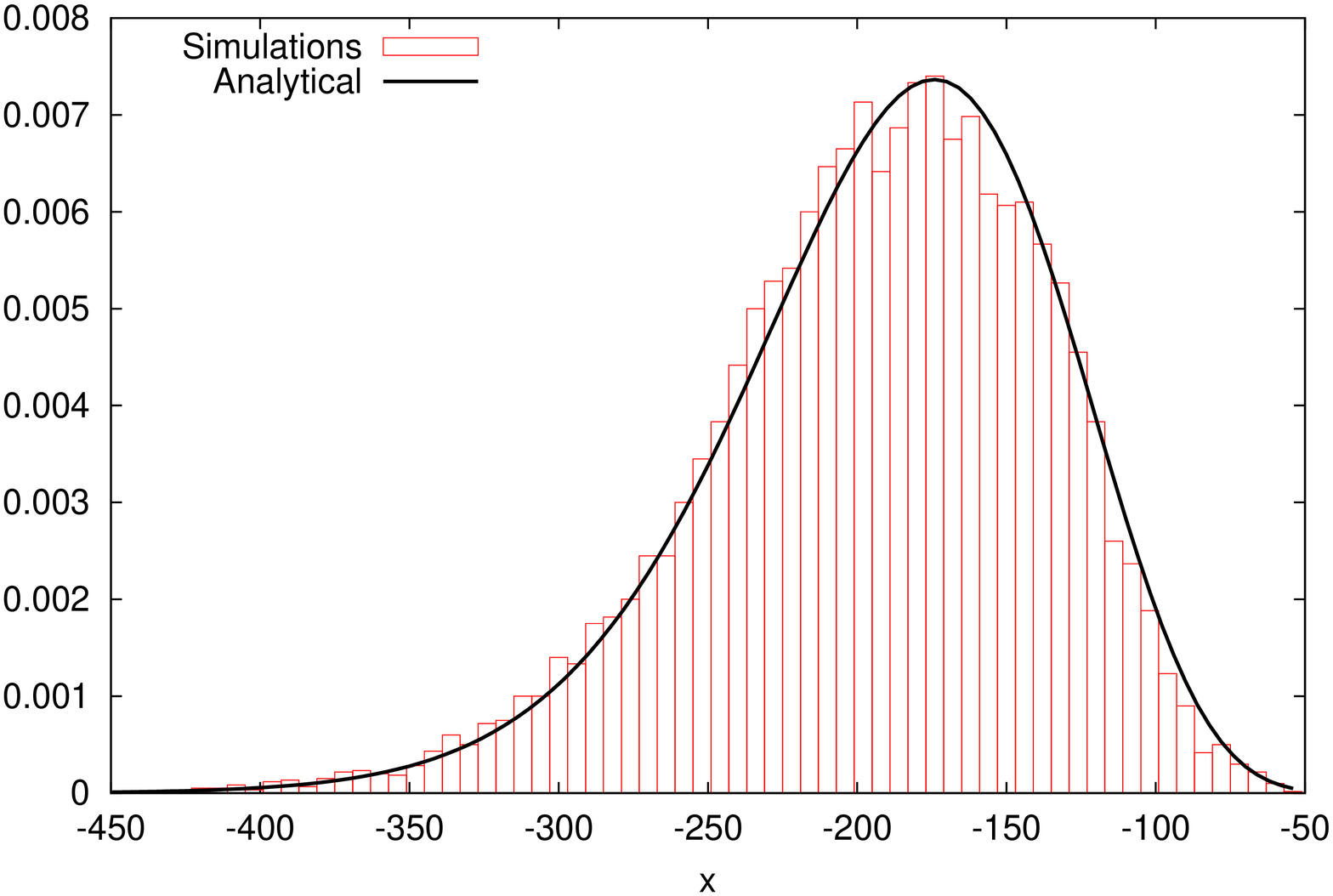}}
	\vskip -3mm
	 \caption{This figure depicts the PDF of some of the coefficients of the form $A^{00}_{ll}$  derived from 15000 simulations. Coefficients with even multipoles ($l$) are left skewed and those with odd multipoles ($l$) are right skewed.}
	  \label{fig5}
\end{figure}

These coefficients are only real, as their imaginary part do not exist. Refer to Appendix \ref{app:BIPOLAR-STATISTICS} for details. In this case, all the terms in summation are independent of each other. In the linear combination there will terms with distinct distribution functions.
Terms with \{$m_1\ne 0, m_1=-m_2$\} are $\chi^2$ distributed with two degrees of freedom and terms with \{$m_1=m_2=0$\} are $\chi^2$ distributed with one degree of freedom.
For the details of the characteristic function of these BipoSH coefficients, refer Appendix \ref{cased}.
\newline The $n^{\textrm{th}}$ order cumulant for $ A^{L0}_{ll}$ can be derived to have the following analytical form,
\begin{eqnarray}
&\tilde K_{n}&=2^{n-1}(C_{l})^{n}(n-1)!\times\left[\left(C^{L0}_{l 0 l 0}\right)^{n}+2\sum\limits_{\substack{m_1\\ \{m_1>0\}}}\left((-1)^{m_1}C^{L0}_{l m_1 l -m_1}\right)^{n}\right] \,. \nonumber
\end{eqnarray}
Moments for these coefficients can be derived given this form of the cumulants (Eq. \ref{eq:moments}). We have shown that the odd moments for these coefficients oscillate between positive and negative values for even and odd multipoles ($l$) respectively. Example of this behavior can be seen in Fig. \ref{fig4} and Fig. \ref{fig5}.
For coefficients with $L\neq0$, mean turns out to be zero but rest of the odd moments
are non-zero which implies that these coefficients have an asymmetric distribution, as seen in Fig. \ref{fig4}.
\newline A special case of these coeffients are of the form $A^{00}_{l l}$, which are equivalent to the CMB angular power spectrum $C_{l}$,
\begin{eqnarray}
A^{00}_{ll}=(-1)^l \sqrt{2l+1}C_l
\end{eqnarray}
The characteristic function for these coefficients has the following form,
\begin{eqnarray}
\varphi_{A^{00}_{l l}}(t)={\left[1-\left({\frac{2 i(-1)^{l}C_{l}t}{\sqrt{2l+1}}}\right)\right]^{-(2l +1)/2}}\,.
\end{eqnarray}
The Fourier transform of the characteristic function yields the probability distribution function(PDF), which is this case is a $\chi^2$ distribution.
For even values of  multipole ($l$) the PDF has the form,
\begin{eqnarray}
f(x,k)=\left\{\begin{array}{cl}
	\frac{1}{2^{k/2}a^{k/2} \Gamma(k/2)} x^{\frac{k}{2}-1}\exp({\frac{-x}{2a}}) &  x\geq0  \\ \\ 0, & \textrm{otherwise}
	   \end{array}\right.
\end{eqnarray}
and for odd values of multipole ($l$), it is $f^{*}(-x,k)$.
Here $a= \frac{C_{l}}{\sqrt{2 l +1}}$ is the related to the isotropic power at multipole $l$ and $k=2l+1$ is number of degrees of freedom of the $\chi^2$ distribution.
The distribution function for these coefficients is asymmetric. The mean for these coefficients are non-vanishing as seen in Fig. \ref{fig5},
this is expected since these are the only non-vanishing coefficients under statistical isotropy.
\subsection*{Case B: Bipolar coefficient with $l_1 \ne l_2, M= 0$}
\begin{figure}[!htbp]
	\centering
    \subfigure[Std. Dev. ($\sigma$)]{\label{fig:edge-a}  \includegraphics[height=5.8cm,width=4cm,angle=-90]{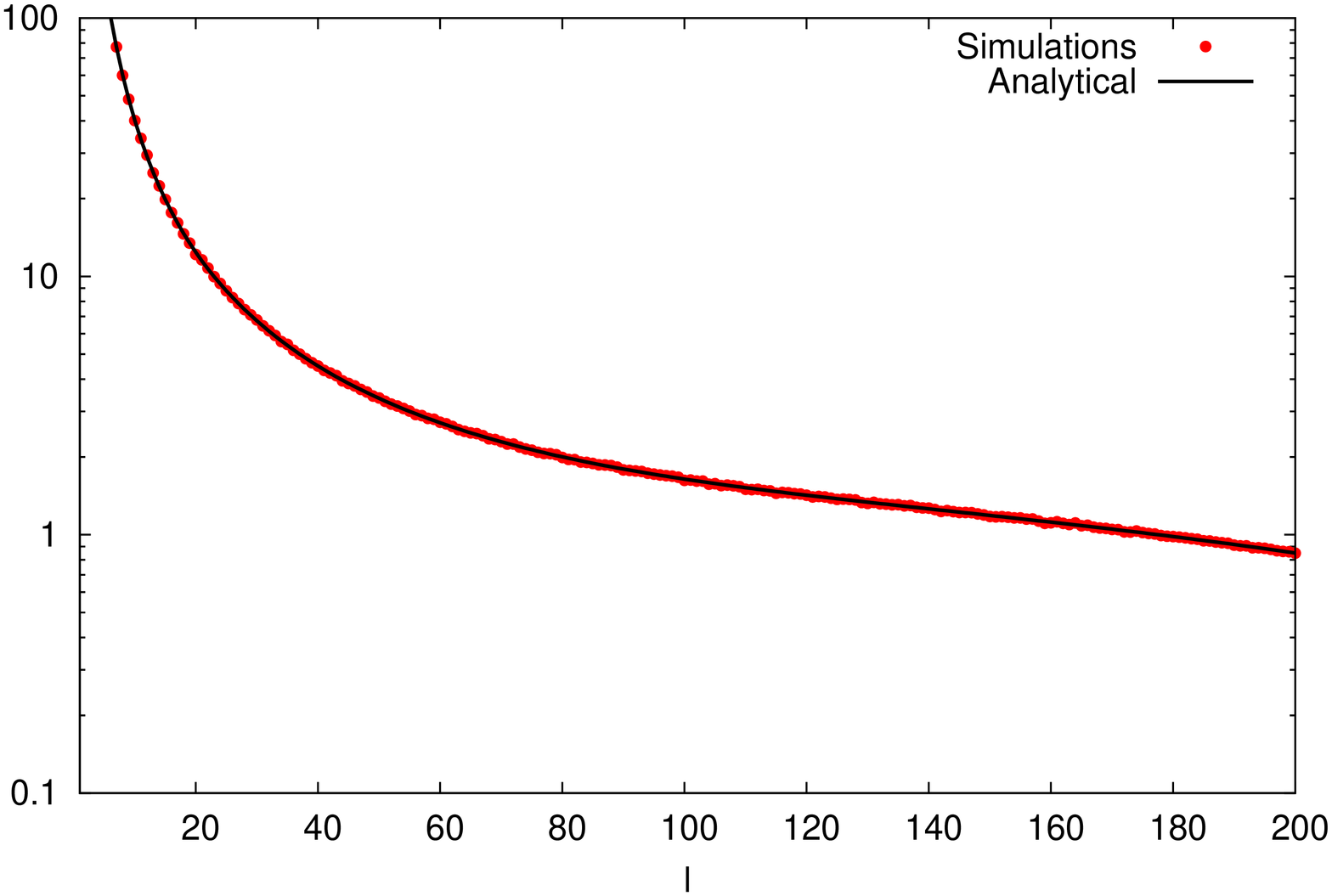}}
	\subfigure[Kurtosis] {\label{fig:edge-a} \includegraphics[height=5.8cm,width=4cm,angle=-90]{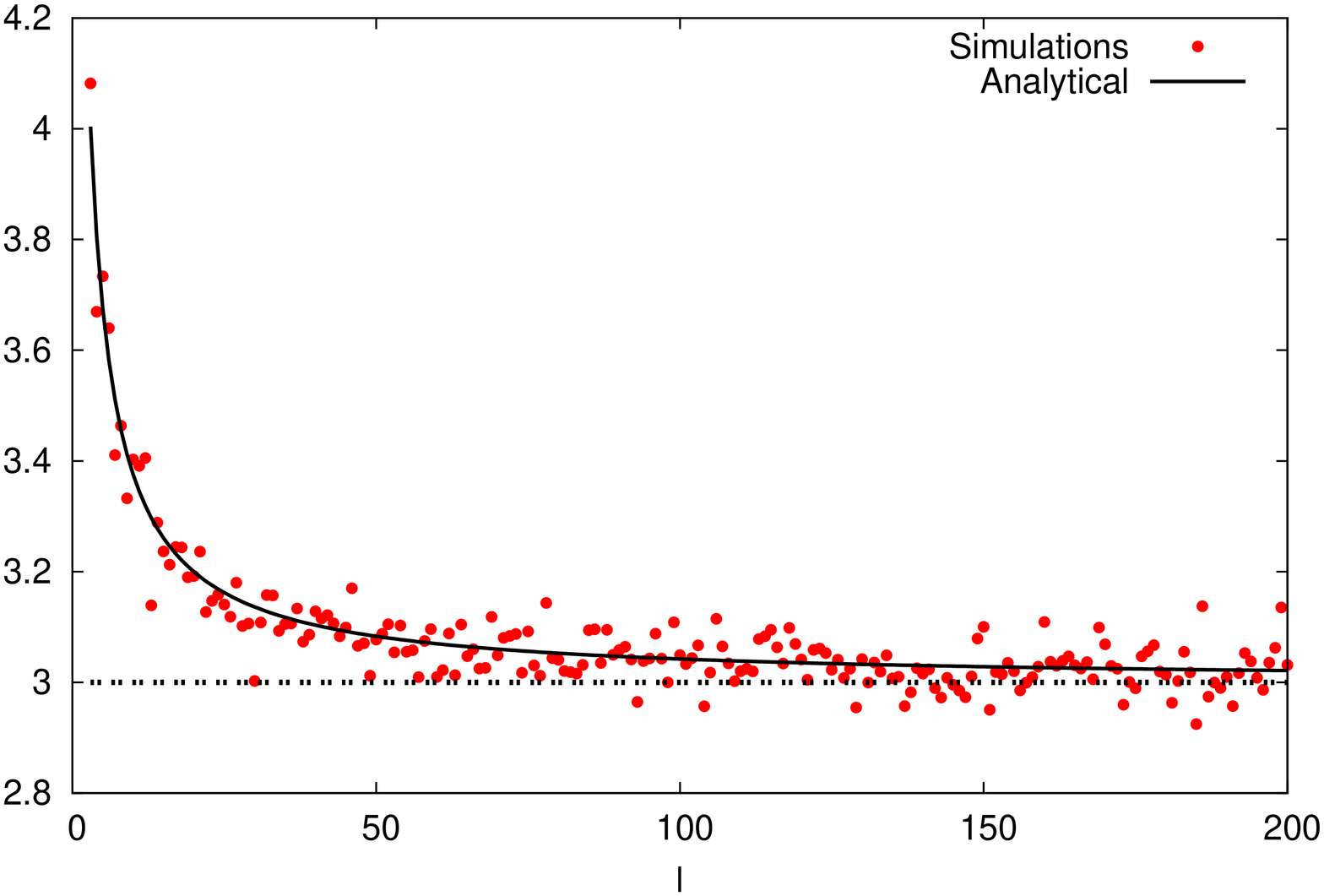}}
    \subfigure[Skewness]{\label{fig:edge-a}  \includegraphics[height=5.8cm,width=4cm,angle=-90]{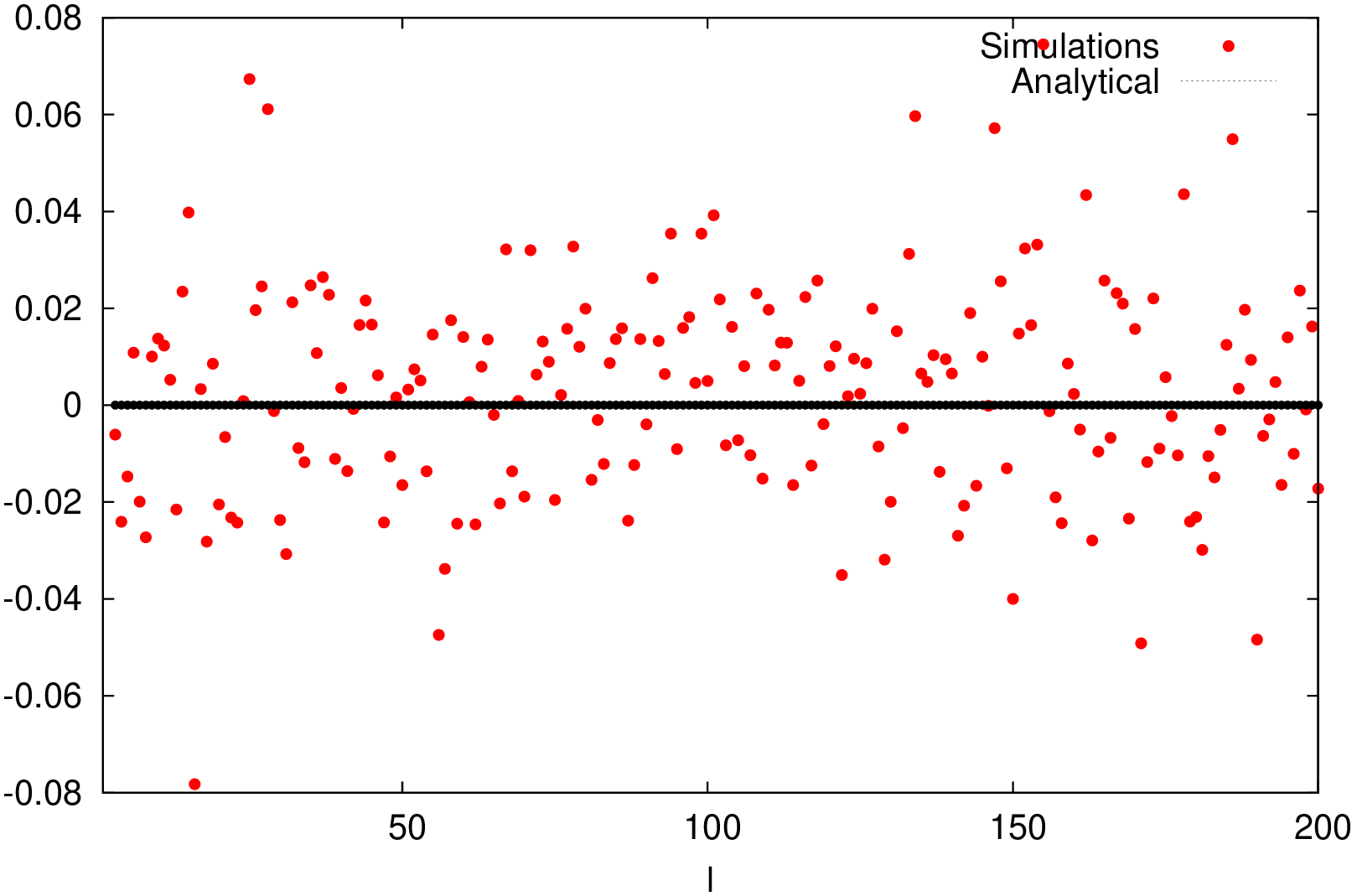}}
	\subfigure[$5^{th}$ Moment] {\label{fig:edge-a} \includegraphics[height=5.8cm,width=4cm,angle=-90]{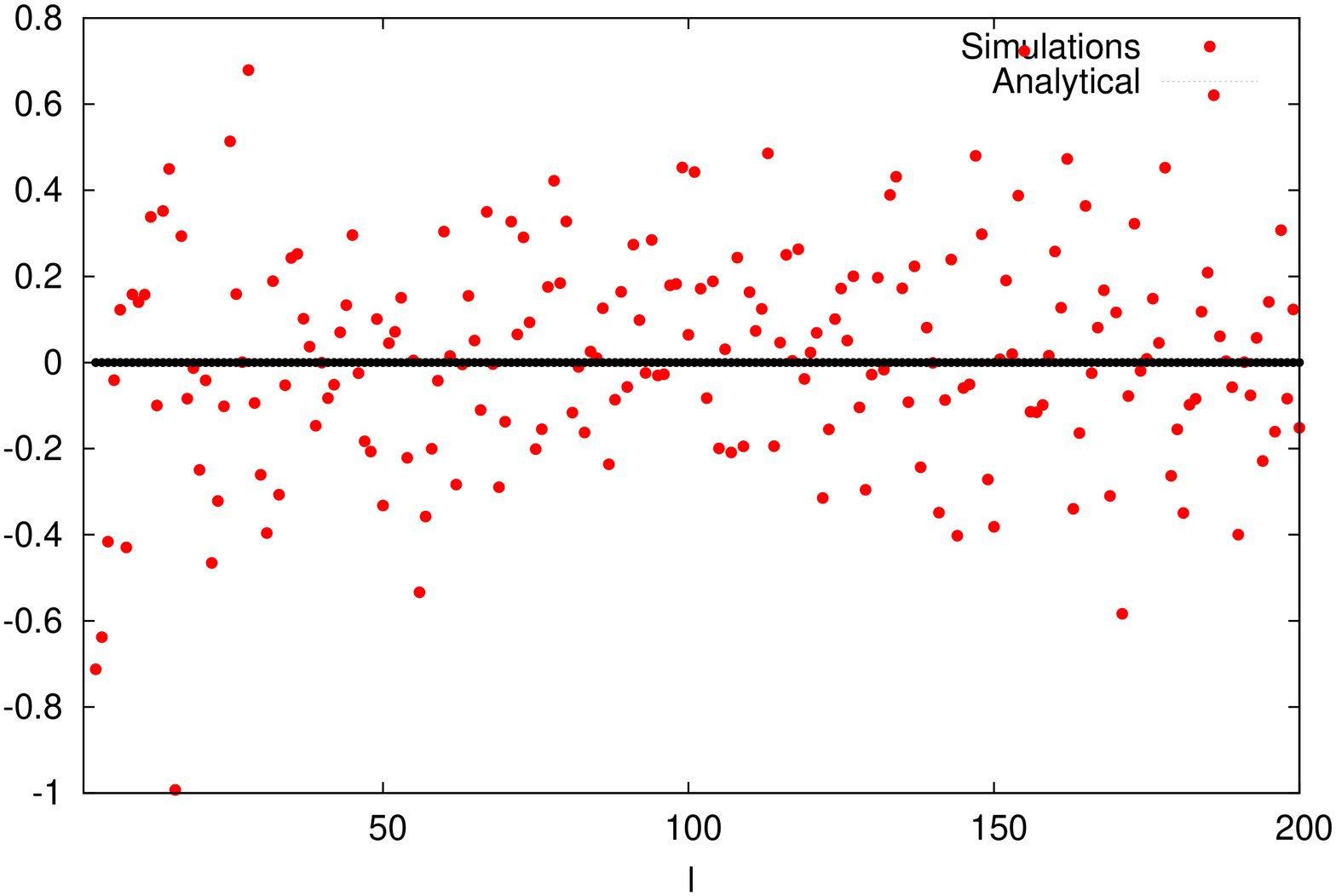}}
	 \caption{Standard deviation, skewness, kurtosis and $5^{th}$ moment of real part of $A^{20}_{l l+2}$, from 15000 simulations. These coefficients have a symmetric PDF. The kurtosis of these coefficients approach that of a Gaussian for high multipoles.} 
	  \label{fig3}
\end{figure}

Even in this case, all the terms in summation are independent of each other. 
In the linear combination there will terms with distinct distribution functions. 
Terms with \{$m_1\ne 0, m_1=-m_2$\} are Laplace distributed and terms with \{$m_1=m_2=0$\} are modified Bessel of second kind distributed. 
The details of the characteristic function of real and imaginary parts of these BipoSH coefficients can be found in Appendix \ref{casec}.
Only even ordered cumulants exist for these coefficients,
\begin{eqnarray}\label{eq:c3}
&&\tilde K_{(n=even)}=(n-1)!(C_{l_1}C_{l_2})^{n/2} \times \Bigg[\sum\limits_{\substack{ m_1\ne0, m_2\ne 0 \\m_1=-m_2}}2^{1-n}\left(C^{LM}_{l_1 m_1 l_2 m_2}\right)^{n} + (C^{LM}_{l_1 0 l_2 0})^{n}\Bigg] \,. \nonumber
\end{eqnarray}
Note that imaginary part of these coefficients will not have any contribution from the second term in above expression for cumulants. Refer Appendix \ref{casec} for details.
Moments of distribution of these coefficients can be obtained given the above form for the cumulants(Eq. \ref{eq:moments}).
These coefficients have symmetric PDF, as evident from Fig. \ref{fig3}.
\subsection*{Case C: Bipolar coefficient with $l_1 = l_2=l, M\ne 0$}
\begin{figure}[!ht]
	\centering
    \subfigure[Std. Dev. ($\sigma$)]{\label{fig:edge-a}  \includegraphics[height=5.8cm,width=4cm,angle=-90]{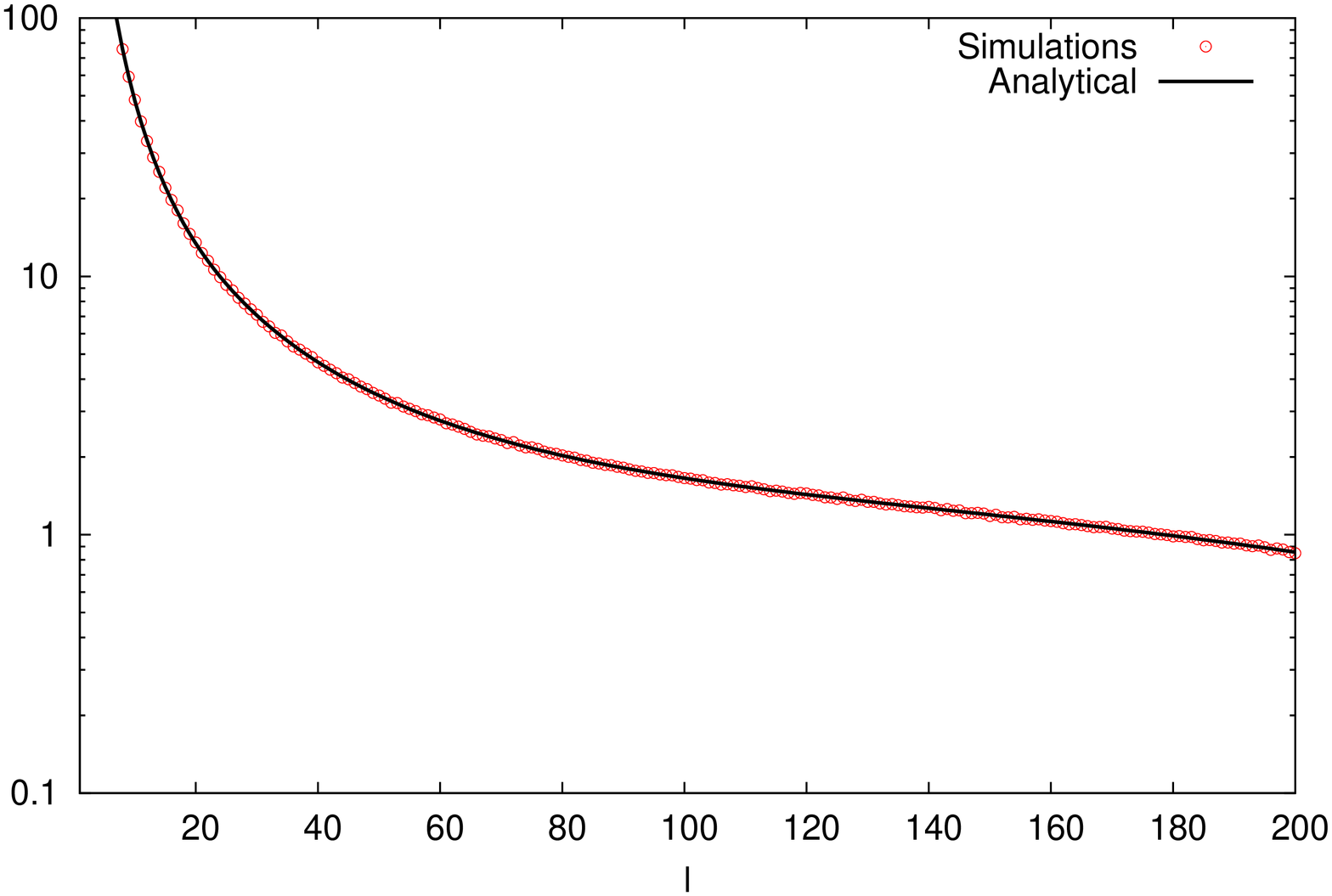}}
	\subfigure[Kurtosis] {\label{fig:edge-a} \includegraphics[height=5.8cm,width=4cm,angle=-90]{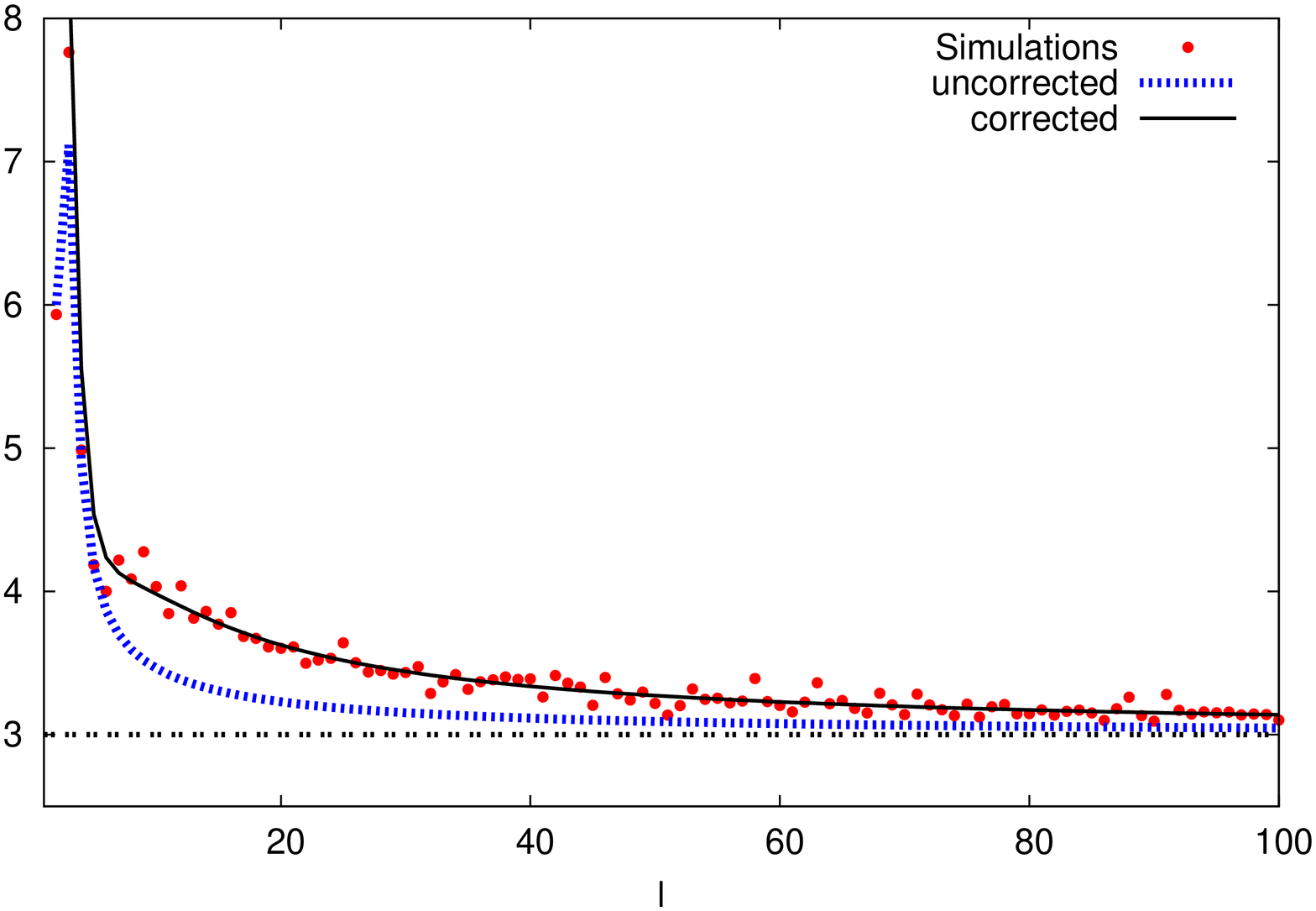}}
	 \caption{Standard deviation and kurtosis of real part of $A^{43}_{l l}$ derived from 15000 simulations. The difference between corrected and uncorrected analytical moments is prominent at low values of multipole ($l$). The corrected kurtosis can be seen to be in good agreement with the simulation resutls. } 
	  \label{fig2}
\end{figure}

We first calculate the moments of distribution for these coefficients using the characteristic function method assuming that all terms in the linear combination are independent. 
In the linear combination for these coefficients there appear terms like \{$m_1\ne0, m_2\ne 0$\} which 
are Laplace distributed and terms like
$\{m_1=0, m_2=M\}$,$\{m_1=M, m_2=0\}$ and \{$m_1=m_2$\} which are distributed as modified Bessel function of second kind of zeroth order. 
The details of the characteristic function for these coefficients can be found in Appendix \ref{caseb}.
It is observed that only even ordered cumulants exist implying that the distribution of these coefficients is symmetric,
\begin{eqnarray}\label{eq:c2}
&&\tilde K_{(n=even)} = 
(n-1)! C_{l}^{n} \times \Bigg[\sum\limits_{\substack{m_1\ne0, m_2\ne0\\m_1>m_2}}2(C^{LM}_{l m_1 l m_2 })^{n}+(C^{LM}_{l m_1 l m_2})^{n}\delta_{m_1 m_2} \nonumber+\sum\limits_{\substack{m_1\vee m_2=0}}
(\sqrt{2}C^{LM}_{l m_1 l m_2 })^{n}\Bigg] \,.
\end{eqnarray}
Note that imaginary part of these coefficients will not have any contribution from the last term in above expression for cumulants. Refer Appendix \ref{caseb} for details.
Moments of distribution of these coefficients can be obtained given the above form for the cumulants(Eq. \ref{eq:moments}), see Fig. (\ref{fig2}) for illustration.
The mismatch in simulations and analytically derived moments at low multipoles ($l$) is due to the assumed underlying independence of the terms contributing to the linear combination, which does not hold true for this case. Hence the characteristic function approach does not fully characterize the statistics of these coefficients. The moments calculated using the characteristic function method need to be supplemented with correction terms, which account for the higher order correlations. Refer to Appendix \ref{correction} for details. However it is found that for certain coefficients the terms involved in the linear combination are all independent and the correction term goes to zero. $\tilde \mu $  are moments calculated using the characteristic function method and $\bar \mu $ are the corrected moments.
\begin{eqnarray}
\bar\mu_n = \tilde \mu_n + \textrm{correction}\,.
\end{eqnarray}
We show that variance will not have any corrections due the fact that the terms are linearly uncorrelated. However kurtosis does have a correction term as seen in Fig. {\ref{fig2}}.
\subsection*{Case D: Bipolar coefficient with $l_1\ne l_2, M\ne 0$}
Similar to the previous case, we begin by finding the moments of distribution for these coefficients using the characteristic function method assuming that all terms in the linear combination are independent. 
In the linear combination for these coefficients there appear terms with \{$m_1\ne0, m_2\ne 0$\} which
are Laplace distributed and terms with
$\{m_1=0, m_2=M\}$, $\{m_1=M, m_2=0\}$ which have modified Bessel function of second kind distribution. 
The details of the characteristic function for these coefficients can be found in Appendix \ref{casea}.
Even for these coefficients it is found that only even ordered cumulants exist implying that their PDF is symmetric.
\begin{eqnarray}\label{eq:c1}
&&\tilde K_{(n=even)}=
(n-1)!(C_{l_1}C_{l_2})^{n/2}\times \Bigg[\sum\limits_{\substack{m_1\ne0, m_2\ne0}}2^{1-n}(C^{LM}_{l_1 m_1 l_2 m_2 })^{n} +\sum\limits_{\substack{m_1\vee m_2=0}} (\sqrt{2})^{-n}(C^{LM}_{l_1 m_1 l_2 m_2 })^{n}\Bigg] \,.
\end{eqnarray}
Note that imaginary part of these coefficients will not have any contribution from the last term in above expression for cumulants. Refer Appendix \ref{casea} for details.
\begin{figure}[!htbp]
	\centering
    \subfigure[Std. Dev. ($\sigma$)]{\label{fig:edge-a}  \includegraphics[height=5.8cm,width=4cm,angle=-90]{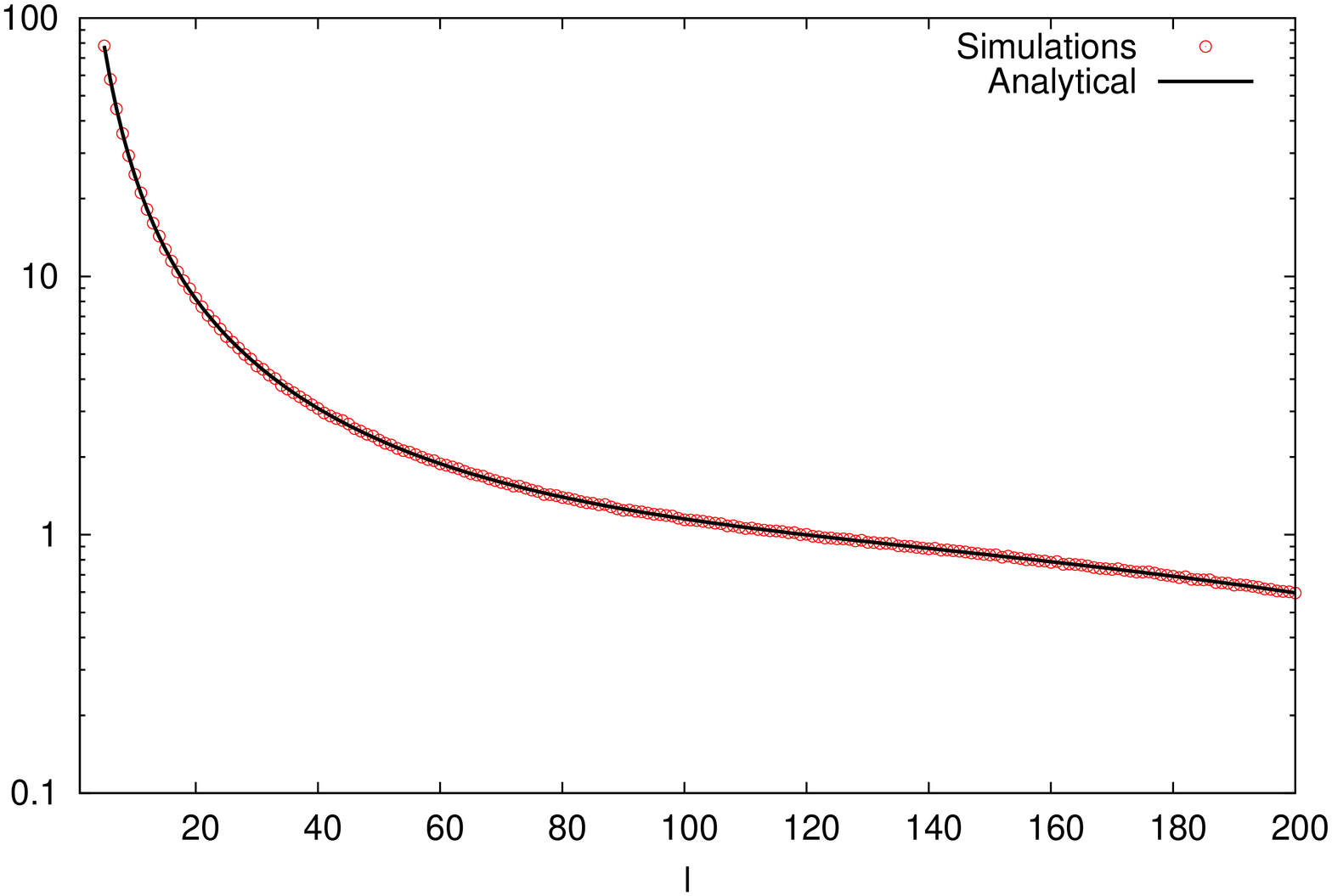}}
	\subfigure[Kurtosis] {\label{fig:edge-a} \includegraphics[height=5.8cm,width=4cm,angle=-90]{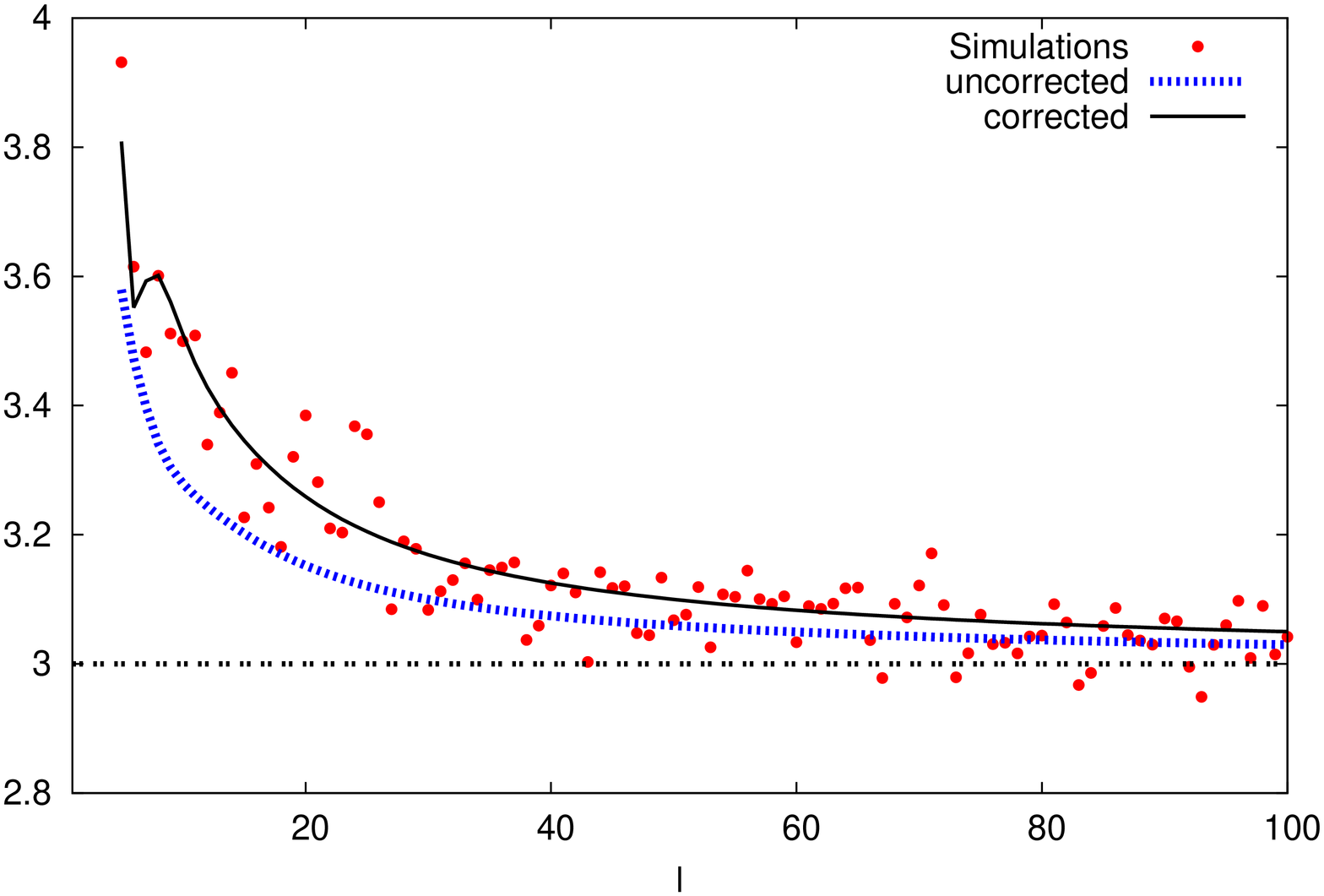}}
	 \caption{Standard deviation and kurtosis of real part of $A^{10~6}_{l l+4}$ derived from 15000 simulations. The difference between corrected and uncorrected analytical moments is prominent at low values of multipole ($l$). The corrected kurtosis can be seen to be in agreement with the simulation results.}
	  \label{fig1}
\end{figure}
Just like in the previous case, the moments calculated using the characteristic function method are supplemented with correction terms which account for the non-linear correlations, see Fig. \ref{fig1} for illustration.
\vskip 1cm
To quantify the agreement between simulations and the analytically derived results we calculate the mean square difference. The closeness of fit is seen (Fig. \ref{chi2}) to be inversely proportional to the number of simulations. We observe that beyond 10000 simulations good convergence is achieved hence we go up to 15000 simulations to derive all our results.
\begin{figure}[!htb]
\centering
\includegraphics[height=8cm,width=4.5cm,angle=-90]{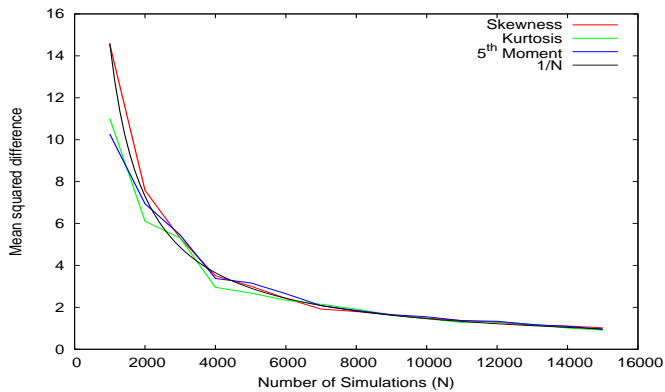}
\vskip 0.4 cm
\caption{Variation of mean squared difference with number of simulations studied for skewness, kurtosis and $5^{th}$ moment of the bipolar coefficients. The mean squared difference evaluated for each moment is multiplied with an arbitrary constant to bring them to the same scale. The mean squared difference for each moment is found to be inversely proportional to the number of simulations.}
\label{chi2}
\end{figure}
\subsection*{Covariance of Bipolar coefficients}
Under statistical isotropy, we show (using Eq. \ref{eq:SI}) that the covariance takes up the following form,
\begin{eqnarray}
\label{eq:covariance}
\langle A^{LM}_{l_1 l_2}A^{*L^{\prime} M^{\prime} }_{l^{\prime}_1 l^{\prime}_2}\rangle&=&C_{l_1}C_{l_2}\delta_{l_1 l^{\prime}_1}\delta_{l_2 l^{\prime}_2}\delta_{L L^{\prime}}\delta_{M M^{\prime}}+(-1)^{l_1+l_2+L}C_{l_1}C_{l_2}\delta_{l_1 l^{\prime}_2}\delta_{l_2 l^{\prime}_1}\delta_{L L^{\prime}}\delta_{M M^{\prime}} \nonumber \\ &+& \Big( C_{l_1}C_{l^{\prime}_1}(-1)^{l_1+l^{\prime}_1}\sqrt{(2l_1+1) (2l^{\prime}_1+1)}\delta_{l_1 l_2}\delta_{l^{\prime}_1 l^{\prime}_2} \nonumber \times \delta_{L0}\delta_{M0}\delta_{L^{\prime} 0}\delta_{M^{\prime} 0} \Big) \,.
\end{eqnarray}
We find that the bipolar coefficients are linearly uncorrelated. However this is not a sufficient condition for the coefficients to be independent of each other.

\section{Discussion and Conclusions}\label{conc}
The temperature field on the CMB sky is usually believed to be a Gaussian random field. Statistical isotropy which implies rotational invariance of two point correlation
function is an assumption in cosmology and needs to be rigorously tested.
Estimators can be constructed aiming at the various kinds of statistical isotropy violations \cite{DH-AL}. Knowing the PDF of these estimators gives a much better handle
on assessing the significance of any statistical isotropy violation detection.
For example, related analysis has also been carried out to find out the PDF of non-Gaussianity estimators, $f_{nl}$, as the significance of a measurement of this parameter depends on 
knowledge of the full shape of its PDF \cite{TS-MK-BW}.
The two point correlation function is used as a measure of statistics of a Gaussian random field.
Correlation function is most generally expanded  in the BipoSH basis. The coefficients of expansion in this basis encode all  the symmetries of the 
correlation function. In this paper we derive the statistical coefficients of these coefficients of expansion. A quantitative understanding of the statistics of these 
coefficients is important, as signal of isotropy violation are being searched for in CMB data in these coefficients.\\
\indent The strategy has been to calculate the characteristic function for these coefficients and then arrive at the cumulants. These cumulants can be easily translated
 to yield the moments of distribution of the coefficients of expansion.  This strategy works perfectly well when the terms involved in the expansion of the BipoSH 
coefficients are independent of each other. However we notice that for a certain set of BipoSH coefficients the characteristic function approach works only partially. 
This is due to the presence of non-linearly correlated terms in the expansion of BipoSH coefficients. In these cases we give a prescription to account for the 
contribution of these non-linear correlations to the moments of the distribution. In this paper we restrict to calculating the correction to the moments only upto 
kurtosis, as calculations for the correction for higher order moments become increasingly tedious, however the general prescription would work.\\
\indent The BipoSH coefficients of the form $A^{00}_{l l}$  are directly related to the CMB angular power spectrum. As expected these coefficients are shown to have a 
$\chi^2$ distribution with ($2l+1$) degrees of freedom using the characteristic function method. For rest of the BipoSH coefficients we provide analytical expressions 
for moments up to any arbitrary order. We find that BipoSH coefficients of the form $A^{L0}_{l l}$ have an asymmetric distribution. The remaining BipoSH coefficients 
are shown to have a symmetric distribution. The BipoSH coefficients of the form $A^{LM}_{l_1 l_2}$ ($M\neq0$) comprise  of terms with non-linear correlations amongst 
them, due to which the analytical moments derived from characteristic function method need to be supplemented with a correction term. We give a general prescription to 
derive the corrections due to the presence of these interdependent terms. We explicitly calculate these correction terms only up to kurtosis.  All these results are 
tested against extensive simulations.
\indent Isotropy violation signal are being sought after in the BipoSH representation of CMB maps. A thorough understanding of the statistics of these coefficients is 
extremely crucial to assess the significance any statistical isotropy violation measurement. In the recent past, WMAP7 team claimed detection of isotropy violation in 
V-Band and W-Band maps. This isotropy violating signal was found in the BipoSH coefficients $A^{20}_{ll}$ and $A^{20}_{ll+2}$. The PDF of these coefficients 
significantly deviate from being Gaussian, particularly, at low multipoles.  The BipoSH coefficients $A^{20}_{ll+2}$ are found to have a symmetric PDF.  

Interestingly in our study we find that the BipoSH coefficients $A^{20}_{ll}$ have an asymmetric PDF, with even multipoles ($l$) being positively skewed and the odd 
multipoles ($l$) being negatively skewed.
The WMAP team uses band power averaged BipoSH coefficients to reduce noise. We find that for full sky and isotropic CMB maps, 
this averaging results in reduced skewness for these coefficients. With experiments like PLANCK it might be possible to achieve similar signal to noise ratio for 
smaller bin sizes. However our study suggests that reducing the bin size, the skewness of these coefficients might become considerable. We are currently assessing the 
implications of these statistics, which is work in progress.


\section*{Acknowledgments}
We acknowledge IUCAA HPC facility and the use of HEALPix package \cite{hpix}. This work initiated from discussions in the CMB group meetings at IUCAA. We acknowledge 
useful discussion with our colleagues, namely, Tuhin Ghosh, Gaurav Goswami, Moumita Aich and Santanu Das. NJ acknowledges support under UGC-JRF scheme 
(Grant award no. 20-12/2009(II)E.U.-IV). AR acknowledges the Council of Scientific and Industrial Research (CSIR) India for financial support 
(Grant award no. 20-6/2008(II)E.U.-IV). TS acknowledges support from the Swarnajayanti fellowship, DST, India.

\appendix
\begin{widetext}
\section*{Appendices}
\section{Statistics of spherical harmonic coefficients}\label{app:SPHAR-STATISTICS}
The temperature fluctuations in the CMB sky maps denoted by $\Delta T(\hat n) ~\textrm{where}~ \hat n = (\theta,\phi)$, can be decomposed in the following manner,
\begin{equation}\label{eq:sp-har}
\Delta T (\hat n)=\sum _{l=1}^{\infty }\sum _
{m=-l}^{m=+l}a_{lm}Y_{lm}(\hat n )\,.
\end{equation}
where $Y_{lm}(\hat n)$ are the spherical harmonics and  $a_{lm}$ are the spherical harmonic coefficients.
The expansion coefficients can be obtained by taking the inverse tranform of the above equation and can be expressed as,
\begin{equation}\label{eq:alm}
a_{lm} = \int d\Omega_{\hat n} Y^*_{lm}({\hat n})\Delta T ({\hat n})\,.
\end{equation}
The spherical harmonics can be expressed in terms of the Legendre polynomials, 
\begin{equation}
Y_{lm}(\theta ,\phi )= (-1)^m
\sqrt{\frac{(2l+1)(l-m)!}{4\pi(l+m)!}} P^{m}_{l}(\cos \theta) {\rm e}
^{im\phi}\,.
\end{equation}
Spherical harmonic coefficients, $a_{lm}$'s are complex coefficients,
\begin{eqnarray}
a_{lm}=x_{lm}+iy_{lm},
\end{eqnarray}
where $x_{lm}$ and $y_{lm}$ are real and imaginary part of $a_{lm}$ and are statistically independent of each other.

Reality of temperature fluctuations (\ref{eq:sp-har}) guarantees that the following relation holds for the spherical harmonic coefficients,
\begin{eqnarray}\label{eq:sphprop1}
a_{lm}&=&(-1)^{m}a^{*}_{l -m},\nonumber \\
x_{lm} &=&(-1)^{m} x_{l -m}, \nonumber \\
y_{lm} &=&(-1)^{m+1} y_{l -m}\,.
\end{eqnarray}

It is easy to see from the above expressions that when $m=0$, the imaginary part of the expansion coefficient vanishes.

CMB temperature fluctuations resulting from the simplest versions of the inflationary paradigm are Gaussian and statistically isotropic. The statistical isotropy(SI) takes the form of a diagonal covariance matrix in harmonic space,
\begin{equation}\label{eq:SI}
<a_{l_1 m_1}a^{*}_{l_2 m_2}>=C_{l_1}\delta_{l_1 l_2}\delta_{m_1 m_2},
\end{equation}
where $C_{l}$ is the angular power spectrum. For case the of statistical isotropy, the angular power spectrum carries all the information about the Gaussian temperature fluctuations.
\newline
The real and imaginary parts of the coefficient $a_{lm}$, with $m\neq0$, are independent Gaussian random variates with mean zero and variances given by,
\begin{equation}
\sigma^{2} (x_{lm}) = \sigma^{2}(y_{lm})=\frac{1}{2} C_l\,.
\end{equation}
However for the coefficients with $m=0$, the imaginary part vanishes and the real part are Gaussian random variables with mean zero and variance given by, 
\begin{equation}
\sigma^{2} (x_{l0}) =C_l\,.
\end{equation}
\section{Characteristic function approach and applications}\label{app:char_approach}
The characteristic function of any random variable is defined as the fourier transform of its  probability distribution function. 
\begin{eqnarray}
 \varphi_{X}(t)=E[e^{itX}]\quad\quad\quad t\in \Re \,.
\end{eqnarray}
 Using the characteristic function to arrive at the statistics of random variables forms a very powerful tool, as there appear situations in which it is easier to arrive at the characteristic function than the probability distribution function itself. This approach is particularly useful in statistical analysis of linear combinations of independent random variables. Let $Z_n$ be a random variable defined in the following manner,
\begin{eqnarray}\label{eq:linear}
 Z_n=\sum^{n}_{i=1}a_i X_i,
\end{eqnarray}
where $X_1,X_2,...X_n$ are independent random variables which are not necessarily identically distributed and $a_i$'s are constants. The characteristic function for  such a random variable ($Z_n$) is given by,
\begin{eqnarray}\label{eq:charprop}
 \varphi_{Z_n}(t)=\varphi_{X_1}(a_1 t)\varphi_{X_2}(a_2 t)......\varphi_{X_n}(a_n t)\,.
\end{eqnarray}
\indent Now we discuss a few applications of this technique, which are extensively used in our calculations.
\linebreak
\indent Let $X_1$ and $X_2$ be two independent normal variates with zero means and variances $\sigma^{2}_1$ and $\sigma^{2}_2$. The distribution of the product of these random variables($Z=X_1 X_2$) is given by\cite{springer,LC-FC},
\begin{eqnarray}\label{eq:npd}
 f_Z(z)=\frac{K_0(\frac{|z|}{\sigma_1 \sigma_2})}{\pi \sigma_1 \sigma_2},
\end{eqnarray}
 where $K_0$ is the zeroth-order modified Bessel function(normal product distribution function).
The characteristic function corresponding to the above distribution function is given by \cite{FM},
\begin{eqnarray}
\varphi_{Z}(t)=\frac{(1/\sigma_1 \sigma_2)}{\left(t^{2}+\frac{1}{\sigma^{2}_1 \sigma^{2}_2} \right)^{1/2}}\,.
\end{eqnarray}

Consider the case of linear combination of two normal product distributed random variates.
If $X_1,Y_1,X_2,Y_2$ are independent Gaussian variates with zero means and variances $\sigma^{2}_1$ for $X_1,Y_1$ and $\sigma^{2}_2$ for $X_2,Y_2$. Then the characteristic function of the random variable $Z=X_1 X_2+Y_1 Y_2$ is given by,
\begin{eqnarray}
 \varphi_{Z}(t)= \frac{1}{\left(1+t^{2}\lambda^{2}\right)},
\end{eqnarray}
where $\lambda=1/\sigma_1 \sigma_2$. \\
The above characteristic function corresponds to that of a Laplace distribution ($Laplace(0,2\lambda^{2})$).
\linebreak
\indent Another interesting application which is of our interest is that of the difference of squares of two Gaussian random variates with zero mean and having the same variance. It is well known that the sum of squares of two Gaussian random variates is $\chi^2$ distributed. The difference however is not $\chi^2$ distributed, instead it has a modified
Bessel function of second kind distribution. We demonstrate this using the characteristic function approach,
If $X$ and $Y$ are two random variables which have normal distribution $N(0,\sigma)$, then $X^{2}$ and $Y^{2}$ are $\chi^2$ distributed and their characteristic function is given by,
\begin{eqnarray}
\varphi(t)= \frac{1}{(1-2i\sigma^{2} t)^{1/2}}\,.
\end{eqnarray}
Using eq.(\ref{eq:charprop}), we obtain the characteristic function for the random variable defined as $Z=X^{2}-Y^{2}$,
\begin{eqnarray}
\varphi_{Z}(t)=\frac{1}{\left[1+(2\sigma^{2} t)^{2}\right]^{1/2}}\,.
\end{eqnarray}
Notice that this characteristic function is that of the  modified Bessel function of second kind distribution with zero order.\\
 
The above illustrated examples are of our particular interest, as they will be used to study the statistics of bipolar spherical harmonic coefficients.
\section{Bipolar statistics} \label{app:BIPOLAR-STATISTICS}

In order to delve the rich source of information which will be provided by future CMB maps, it is important to device methods
to detect, isolate and diagnose various possible causes of departure from statistical isotropy.
In particular, our approach is to look at the statistical behavior of the complex coefficient that arise in bipolar spherical analysis of the CMB two point correlation function.
\begin{eqnarray}
 A^{LM}_{l_1 l_2}=A^{LM^{(R)}}_{l_1 l_2}+i A^{LM^{(I)}}_{l_1 l_2}\,.
\end{eqnarray}
Owing to the reality of the correlation function, the following relation holds between the BipoSH coefficients,
\begin{eqnarray}
A^{*LM}_{l_1 l_2}&=&(-1)^{l_1+l_2-L+M}A^{L -M}_{l_1 l_2} ,\nonumber \\
A^{LM^{(R)}}_{l_1 l_2} &=& (-1)^{l_1+l_2-L+M}A^{L -M^{(R)}}_{l_1 l_2},\nonumber \\
A^{LM^{(I)}}_{l_1 l_2} &=& (-1)^{(l_1+l_2-L+M)+1}A^{L -M^{(I)}}_{l_1 l_2}\,.
\end{eqnarray}
The real and imaginary parts of the BipoSH coefficients can be expressed as linear combinations of the elements of the covariance matrix in harmonic space,
\begin{eqnarray}\label{eq:BipoSH-exp}
 A^{LM^{(R)}}_{l_1 l_2} &=& \sum_{m_1 m_2}(x_{l_1 m_1}x_{l_2 m_2}-y_{l_1 m_1}y_{l_2 m_2})C^{LM}_{l_1 m_1 l_2 m_2},\nonumber \\
A^{LM^{(I)}}_{l_1 l_2} &=& \sum_{m_1 m_2}(y_{l_1 m_1}x_{l_2 m_2}+x_{l_1 m_1}y_{l_2 m_2})C^{LM}_{l_1 m_1 l_2 m_2}\,.
\end{eqnarray}
The indices in the above expression satisfy the following relations $: |l_1-l_2|\le L\le l_1+l_2 \textrm{ and }m_1+m_2=M $, owing to the presence of the Clebsch-Gordon coefficients.\\
The BipoSH coefficients can be classified on the basis of the form of their  characteristic function,
\newline\textbf{Case A}:  $l_1 = l_2, M=0$,
\newline\textbf{Case B}:  $l_1\ne l_2, M = 0$,
\newline\textbf{Case C}:  $l_1 = l_2, M\ne 0$,
\newline\textbf{Case D}:  $l_1\ne l_2, M\ne 0$.

\subsection{Case A: $l_1=l_2, M = 0$}\label{cased}
These BipoSH coefficient can be expressed in terms of the spherical harmonic coeficients of the CMB maps,
\begin{eqnarray}
 A^{L0}_{l_1 l_1}&=&\sum_{\substack{m_1 m_2\\ \{(m_1\ne0, m_2\ne0),m_1=-m_2\}}}a_{l_1 m_1}a_{l_1
m_2}C^{L0}_{l_1 m_1 l_1 m_2} \nonumber \\ &+ &\sum_{\substack{m_1 m_2 \\ \{m_1=m_2=0\}}}a_{l_1 m_1}a_{l_1
m_2}C^{L0}_{l_1 m_1 l_1 m_2}\,.
\end{eqnarray}
We have divided the expansion of the BipoSH coefficients into two type of terms. Terms with $\{(m_1\ne0, m_2\ne0),m_1=-m_2\}$ and the terms where both $m_1$ and $m_2$ are zero $\{m_1=m_2=0\}$. This is done because of the fact that each of these terms have distinct distributions. The imaginary part of these coefficients vanish owing to the reality of the correlation function. \\
The real part of these coefficients is given by (Eq. \ref{eq:BipoSH-exp}), 
\begin{eqnarray}
 A^{L0^{(R)}}_{l_1 l_1}&= \substack{\sum\\ {m_1 (m_1>0)}}(-1)^{m_1} 2 (x^{2}_{l_1 m_1}+y^{2}_{l_1 m_1})C^{L0}_{l_1 m_1 l_1 -m_1}+ x^{2}_{l_1 0}C^{L0}_{l_1 0 l_1 0}\,.
\end{eqnarray}
To arrive at the moments of these BipoSH coefficients, one needs the characteristic function of each term in the summation.  The first term in the above expression has a $\chi^2$ distribution with two degrees of freedom. Its characteristic function has the following form (refer Appendix \ref{app:char_approach}),
\begin{eqnarray}
 \varphi_{Z}(t)= \frac{1}{\left[1-\left({2i(-1)^{m_1}C^{L0}_{l_1 m_1 l_1 -m_1}C_{l_1}t}\right)\right]}\,.
\end{eqnarray}
The second term is $\chi^2$ distributed with one degree of freedom and its characteristic function has the following form (refer Appendix \ref{app:char_approach}),
\begin{eqnarray}
 \varphi_{Z}(t)= \frac{1}{\left[1-\left({2 i C^{L0}_{l_1 0 l_1 0}C_{l_1}t}\right)\right]^{1/2}}\,.
\end{eqnarray}
Hence, the characteristic function of these BipoSH coefficients is given by,
\begin{eqnarray}\label{eq:MGF4}
\varphi_{A^{L0^{(R)}}_{l_1 l_1}}(t)&=&\left[\prod_{\substack{m_1\\ \{m_1\ne0\}}}\frac{1}{\left[1-\left({2i(-1)^{m_1}C^{L0}_{l_1 m_1 l_1 -m_1}C_{l_1}t}\right)\right]}\right] \times \left[ \frac{1}{\left[1-\left({2 i C^{L0}_{l_1 0 l_1 0}C_{l_1}t}\right)\right]^{1/2}}\right]\,.
\end{eqnarray}
\subsection{Case B: $l_1\ne l_2, M = 0$}\label{casec}
The difference in the expansion in this case and the case above is that here $l_1 \neq l_2$.
\begin{eqnarray}
 A^{LM}_{l_1 l_2}&=&\sum_{\substack{m_1 m_2\\ \{(m_1\ne0, m_2\ne0),m_1=-m_2\}}}a_{l_1 m_1}a_{l_2
m_2}C^{LM}_{l_1 m_1 l_2 m_2}+\sum_{\substack{m_1 m_2 \\ \{m_1=m_2=0\}}}a_{l_1 m_1}a_{l_2
m_2}C^{LM}_{l_1 m_1 l_2 m_2}\,.
\end{eqnarray}
\indent The real and imaginary part of these coefficients are given by the following expressions (Eq. \ref{eq:BipoSH-exp}),
\begin{eqnarray}
 A^{L0^{(R)}}_{l_1 l_2} &=& \sum_{\substack{m_1 m_2\\ \{(m_1\ne0, m_2\ne0),m_1=-m_2\}}}(x_{l_1 m_1}x_{l_2 m_2}-y_{l_1 m_1}y_{l_2 m_2})C^{L0}_{l_1 m_1 l_2 m_2}+x_{l_1 0}x_{l_2 0}C^{L0}_{l_1 0 l_2 0},\nonumber \\
A^{L0^{(I)}}_{l_1 l_2} &=& \sum_{\substack{m_1 m_2\\ \{(m_1\ne0, m_2\ne0),m_1=-m_2\}}}(y_{l_1 m_1}x_{l_2 m_2}+x_{l_1 m_1}y_{l_2 m_2})C^{L0}_{l_1 m_1 l_2 m_2}\,.
\end{eqnarray}
Note that the imaginary part of these coefficients does not vanish. \\
The first term in the expansion for $A^{L0^{(R)}}_{l_1 l_2}$ and  $A^{L0^{(I)}}_{l_1 l_2}$ is Laplace distributed with characteristic function given by (refer Appendix \ref{app:char_approach}),
\begin{eqnarray}
 \varphi_{Z}(t)= \frac{2}{\pi(4+(C^{L0}_{l_1 m_1 l_2 m_2} t \sqrt{C_{l_1}C_{l_2}})^{2})}
\end{eqnarray}
and the second term in the expansion for $A^{L0^{(R)}}_{l_1 l_2}$ has modified Bessel function of second kind distribution with the following characteristic function (refer Appendix \ref{app:char_approach}),
\begin{eqnarray}
 \varphi_{Z}(t)= \frac{1}{\sqrt{\pi}\sqrt{2+( C^{L0}_{l_1 m_1 l_2 m_2} t \sqrt{C_{l_1}C_{l_2}} )^{2}}}\,.
\end{eqnarray}
Hence, the characteristic function for the real part of these BipoSH coefficients is given by,
\begin{eqnarray}\label{eq:MGF3}
\varphi_{A^{LM^{(R)}}_{l_1 l_2}}(t)&=&\left[\prod_{\substack{m_1 m_2\\ \{(m_1\ne0, m_2\ne0),m_1=-m_2\}}}\frac{2}{\pi(4+(C^{L0}_{l_1 m_1 l_2 m_2} t \sqrt{C_{l_1}C_{l_2}})^{2})}\right] \nonumber\\ &\times & 
\left[\prod_{\substack{m_1 m_2\\ \{(m_1=0, m_2=0)\}}}\frac{1}{\sqrt{\pi}\sqrt{2+( C^{L0}_{l_1 m_1 l_2 m_2} t \sqrt{C_{l_1}C_{l_2}} )^{2}}}\right]
\end{eqnarray} 
and the characteristic function for the imaginary part of these BipoSH coefficients is given by,
\begin{eqnarray}\label{eq:MGFI3}
\varphi_{A^{LM^{(I)}}_{l_1 l_2}}(t)=\left[\prod_{\substack{m_1 m_2\\ \{(m_1\ne0, m_2\ne0),m_1=-m_2\}}}\frac{2}{\pi(4+(C^{L0}_{l_1 m_1 l_2 m_2} t \sqrt{C_{l_1}C_{l_2}})^{2})}\right]\,.
\end{eqnarray} 
\subsection{Case C: $l_1 = l_2, M\ne 0$}\label{caseb}
The expansion of these BipoSH coefficients is split into three parts depending upon the form of the characteristic function of each of the terms,
\begin{eqnarray}
 A^{LM}_{l_1 l_1}=\sum_{\substack{m_1 m_2\\ \{(m_1\ne0, m_2\ne0), m_1>m_2\}}}2 a_{l_1 m_1}a_{l_1
m_2}C^{LM}_{l_1 m_1 l_1 m_2}+\nonumber\\ \sum_{\substack{m_1 m_2 \\ \{(m_1\vee m_2)=0, m_1>m_2\}}}2 a_{l_1 m_1}a_{l_1
m_2}C^{LM}_{l_1 m_1 l_1 m_2}+ \sum_{\substack{m_1 m_2 \\ \{m_1=m_2\}}}a_{l_1 m_1}a_{l_1 m_2}C^{LM}_{l_1 m_1 l_1 m_2}\,.
\end{eqnarray}
The real and imaginary parts of these bipolar coefficients are given by,
\begin{eqnarray}
 A^{LM^{(R)}}_{l_1 l_1}=\sum_{\substack{m_1 m_2\\ \{(m_1\ne0, m_2\ne0), m_1>m_2\}}}2 (x_{l_1 m_1}x_{l_1 m_2}-y_{l_1 m_1}y_{l_1 m_2})C^{LM}_{l_1 m_1 l_1 m_2}+
\nonumber\\ \sum_{\substack{m_1 m_2 \\ \{(m_1\vee m_2)=0, m_1>m_2\}}}2 x_{l_1 m_1}x_{l_1 m_2}C^{LM}_{l_1 m_1 l_1 m_2}+ 
\sum_{\substack{m_1 m_2 \\ \{m_1=m_2\}}}(x_{l_1 m_1}x_{l_1 m_2}-y_{l_1 m_1}y_{l_1 m_2})C^{LM}_{l_1 m_1 l_1 m_2},
\nonumber
\end{eqnarray}
\begin{eqnarray}
 A^{LM^{(I)}}_{l_1 l_1} = \sum_{\substack{m_1 m_2\\ \{(m_1\ne0, m_2\ne0), m_1>m_2\}}}2(y_{l_1 m_1}x_{l_1 m_2}+x_{l_1 m_1}y_{l_1 m_2})C^{LM}_{l_1 m_1 l_1 m_2}+
\nonumber\\ 
\sum_{\substack{m_1 m_2 \\ \{m_1=m_2\}}}(y_{l_1 m_1}x_{l_1 m_2}+x_{l_1 m_1}y_{l_1 m_2})C^{LM}_{l_1 m_1 l_1 m_2}\,.
\end{eqnarray}
The first term in the expansion for $A^{L0^{(R)}}_{l_1 l_2}$ and  $A^{L0^{(I)}}_{l_1 l_2}$ is Laplace distributed with characteristic function given by (refer Appendix \ref{app:char_approach}),
\begin{eqnarray}
 \varphi_{Z}(t)= \frac{2}{\pi(4+(2 C^{LM}_{l_1 m_1 l_1 m_2} C_{l_1}t)^{2})}\,.
\end{eqnarray}
The second term in the expansion for $A^{L0^{(R)}}_{l_1 l_2}$ has a modified Bessel function of second kind distribution. It has the following characteristic function (refer Appendix \ref{app:char_approach}),
\begin{eqnarray}
 \varphi_{Z}(t)= \frac{1}{\sqrt{\pi}\sqrt{2+(2 C^{LM}_{l_1 m_1 l_1 m_2} C_{l_1}t )^{2}}}\,.
\end{eqnarray}
The last terms in the expansion for $A^{L0^{(R)}}_{l_1 l_2}$ and  $A^{L0^{(I)}}_{l_1 l_2}$ have
a modified bessel function of second kind distribution and the corresponding characteristic function is given by  (refer Appendix \ref{app:char_approach}),
\begin{eqnarray}
 \varphi_{Z}(t)= \frac{1}{\sqrt{2\pi}\sqrt{1+(C^{LM}_{l_1 m_1 l_1 m_2}C_{l_1}t )^{2}}}\,.
\end{eqnarray}
The characteristic function for the real part of BipoSH coefficients can now be easily derived to have the following form,
\begin{eqnarray}\label{eq:MGF2}
\varphi_{A^{LM^{(R)}}_{l_1 l_1}}(t)&=&\left[\prod_{\substack{m_1 m_2\\ \{(m_1\ne0, m_2\ne0), m_1>m_2\}}}\frac{2}{\pi(4+(2 C^{LM}_{l_1 m_1 l_1 m_2} C_{l_1}t)^{2})}\right]\nonumber\\ &\times &
\left[\prod_{\substack{m_1 m_2 \\ \{(m_1\vee m_2)=0, m_1>m_2\}}}\frac{1}{\sqrt{\pi}\sqrt{2+(2 C^{LM}_{l_1 m_1 l_1 m_2} C_{l_1}t )^{2}}}\right] \nonumber \\ &\times&
\left[ \prod_{\substack{m_1 m_2 \\ \{m_1=m_2\}}}\frac{1}{\sqrt{2\pi}\sqrt{1+(C^{LM}_{l_1 m_1 l_1 m_2}C_{l_1}t )^{2}}}\right]\,.
\end{eqnarray} 
and the imaginary part of the BipoSH coefficients can be derived to have the following form,
\begin{eqnarray}\label{eq:MGFI2}
\varphi_{A^{LM^{(I)}}_{l_1 l_1}}(t)&=&\left[\prod_{\substack{m_1 m_2\\ \{m_1\ne0, m_2\ne0 \}\\\{ m_1>m_2\}}}\frac{2}{\pi(4+(2 C^{LM}_{l_1 m_1 l_1 m_2} C_{l_1}t)^{2})}\right] \times
\left[ \prod_{\substack{m_1 m_2 }}\frac{\delta_{m_1 m_2}}{\sqrt{2\pi}\sqrt{1+(C^{LM}_{l_1 m_1 l_1 m_2}C_{l_1}t )^{2}}}\right]\,.
\end{eqnarray} 
\subsection{Case D: $l_1\ne l_2, M\ne 0$}\label{casea}
The BipoSH coefficients in this case will have the following expansion,
\begin{eqnarray}
 A^{LM}_{l_1 l_2}=\sum_{\substack{m_1 m_2\\ \{m_1\ne0, m_2\ne0\}}}a_{l_1 m_1}a_{l_2
m_2}C^{LM}_{l_1 m_1 l_2 m_2}+\sum_{\substack{m_1 m_2 \\ \{(m_1\vee m_2)=0\}}}a_{l_1 m_1}a_{l_2
m_2}C^{LM}_{l_1 m_1 l_2 m_2}\,.
\end{eqnarray}
The real and imaginary parts of these coefficients can be expressed as (Eq.\ref{eq:BipoSH-exp}),
\begin{eqnarray}
 A^{LM^{(R)}}_{l_1 l_2} &=& \sum_{\substack{m_1 m_2\\ \{m_1\ne0, m_2\ne0\}}}(x_{l_1 m_1}x_{l_2 m_2}-y_{l_1 m_1}y_{l_2 m_2})C^{LM}_{l_1 m_1 l_2 m_2}\nonumber \\ &+&
\sum_{\substack{m_1 m_2 \\ \{(m_1\vee m_2)=0\}}}x_{l_1 m_1}x_{l_2 m_2}C^{LM}_{l_1 m_1 l_2 m_2},\nonumber\\
 A^{LM^{(I)}}_{l_1 l_2} &=& \sum_{\substack{m_1 m_2\\ \{m_1\ne0, m_2\ne0\}}}(y_{l_1 m_1}x_{l_2 m_2}+x_{l_1 m_1}y_{l_2 m_2})C^{LM}_{l_1 m_1 l_2 m_2}\,.
\end{eqnarray} 
The first term in the expansion for $A^{L0^{(R)}}_{l_1 l_2}$ and  $A^{L0^{(I)}}_{l_1 l_2}$ is Laplace distributed and its characteristic function given by (refer Appendix \ref{app:char_approach}),
\begin{eqnarray}
 \varphi_{Z}(t)= \frac{2}{\pi(4+(C^{LM}_{l_1 m_1 l_2 m_2}t\sqrt{C_{l_1}C_{l_2}})^{2})}\,.
\end{eqnarray}
The second term in the expansion for $A^{L0^{(R)}}_{l_1 l_2}$ has a modified Bessel function of second kind distribution. It has the following characteristic function (refer Appendix \ref{app:char_approach}),
\begin{eqnarray}
 \varphi_{Z}(t)= \frac{1}{\sqrt{\pi}\sqrt{2+(C^{LM}_{l_1 0 l_2 m_2}t\sqrt{C_{l_1}C_{l_2}})^{2}}}\,.
\end{eqnarray}
Hence, the characteristic function for the real part of BipoSH coefficients is the product of all the characteristic functions of individual terms and has the following form,
\begin{eqnarray}\label{eq:MGF1}
\varphi_{A^{LM^{(R)}}_{l_1 l_2}}(t)=\left[\prod_{\substack{m_1,m_2 \\ \{m_1\ne0, m_2\ne0\}}}\frac{2}{\pi(4+(C^{LM}_{l_1 m_1 l_2 m_2}t\sqrt{C_{l_1}C_{l_2}})^{2})}\right]\times \nonumber\\
\left[\prod_{\substack{m_1 m_2 \\ \{(m_1\vee m_2)=0\}}}\frac{1}{\sqrt{\pi}\sqrt{2+(C^{LM}_{l_1 m_1 l_2 m_2}t\sqrt{C_{l_1}C_{l_2}})^{2}}}\right]\,.
\end{eqnarray} 
and the characteristic function for the imaginary part of the BipoSH coefficients is given by,
\begin{eqnarray}\label{eq:MGFI1}
\varphi_{A^{LM^{(I)}}_{l_1 l_2}}(t)=\left[\prod_{\substack{m_1,m_2 \\ \{m_1\ne0, m_2\ne0\}}}\frac{2}{\pi(4+(C^{LM}_{l_1 m_1 l_2 m_2}t\sqrt{C_{l_1}C_{l_2}})^{2})}\right]\,.
\end{eqnarray}
\subsection{Covariance of BipoSH coefficients}\label{covariance}
The unbiased estimator of the BipoSH coefficients is given by,
\begin{eqnarray}
A^{LM}_{l_1 l_2}=\sum_{{m_1 m_2}}(-1)^{m_2}a_{l_1 m_1}a^{*}_{l_2 m_2}C^{LM}_{l_1 m_1 l_2 -m_2}\,.
\end{eqnarray}
The covariance of these coefficients is defined in the following manner,
\begin{eqnarray}
\langle A^{LM}_{l_1 l_2}A^{*L^{\prime} M^{\prime} }_{l^{\prime}_1 l^{\prime}_2}\rangle=\left(\sum_{m_1 m_2}\sum_{m^{\prime}_1 m^{\prime}_2}
(-1)^{m_2+m^{\prime}_2}\langle a_{l_1 m_1}a^{*}_{l_2 m_2}
a^{*}_{l^{\prime}_1 m^{\prime}_1}a_{l^{\prime}_2 m^{\prime}_2}\rangle 
C^{LM}_{l_1 m_1 l_2 -m_2}C^{L^{\prime}M^{\prime}}_{l^{\prime}_1 m^{\prime}_1 l^{\prime}_2 -m^{\prime}_2}\right)\,.
\end{eqnarray}
The spherical harmonic coefficients  ($a_{lm}$'s) are Gaussian random variables, hence the four point correlation function appearing in the above equation can be written as,
\begin{eqnarray}\label{eq:realbiposh3}
\langle a_{l_1 m_1}a^{*}_{l_2 m_2}a^{*}_{l^{\prime}_1 m^{\prime}_1}a_{l^{\prime}_2 m^{\prime}_2}\rangle &=&
\langle a_{l_1 m_1}a^{*}_{l_2 m_2}\rangle\langle a^{*}_{l^{\prime}_1 m^{\prime}_1}a_{l^{\prime}_2 m^{\prime}_2}\rangle \nonumber \\
&+&\langle a_{l_1 m_1}a^{*}_{l^{\prime}_1 m^{\prime}_1}\rangle\langle a^{*}_{l_2 m_2}a_{l^{\prime}_2 m^{\prime}_2}\rangle \nonumber \\
&+&\langle a_{l_1 m_1}a_{l^{\prime}_2 m^{\prime}_2}\rangle\langle a^{*}_{l^{\prime}_1 m^{\prime}_1} a^{*}_{l_2 m_2}\rangle\,.
\end{eqnarray}
Under the assumption of statistical isotropy, the covariance of the BipoSH coefficients can be derived to have the following form (Eq.\ref{eq:SI}),
\begin{eqnarray}\label{eq:covariance}
\langle A^{LM}_{l_1 l_2}A^{*L^{\prime} M^{\prime} }_{l^{\prime}_1 l^{\prime}_2}\rangle &=&C_{l_1}C_{l^{\prime}_1}(-1)^{l_1+l^{\prime}_1}
\left[(2l_1+1) (2l^{\prime}_1+1)\right]^{1/2}\delta_{l_1 l_2}\delta_{l^{\prime}_1 l^{\prime}_2}\delta_{L0}\delta_{M0}\delta_{L^{\prime} 0}
\delta_{M^{\prime} 0} \nonumber \\
&+&C_{l_1}C_{l_2}\delta_{l_1 l^{\prime}_1}\delta_{l_2 l^{\prime}_2}\delta_{L L^{\prime}}\delta_{M M^{\prime}} 
+ (-1)^{l_1+l_2+L}C_{l_1}C_{l_2}\delta_{l_1 l^{\prime}_2}\delta_{l_2 l^{\prime}_1}\delta_{L L^{\prime}}\delta_{M M^{\prime}}\,. \nonumber \\
\end{eqnarray}

\subsection{Correction to moments due non-linear correlations.}\label{correction}

Consider a random variable define by,
\begin{equation}
Z=\sum_{i}^{N}X_{i},
\end{equation}
where $X_i$'s are random variable with arbitrary distributions and not necessarily independent and $N$ is total number of terms.

Any arbitrary moment of the distribution of the random variable Z can be expressed as,
\begin{eqnarray}
\langle Z^n \rangle = \langle ( \sum_{i}^{N} X_i )^n \rangle\,.
\end{eqnarray}

In the case where the random variables are all independent of each other, the above expression will acquire this simple form,
\begin{eqnarray}
\langle Z^n \rangle = \sum_{i}^{N} \langle (X_i)^n  \rangle\,.
\end{eqnarray}
However in the case where the random variables are not all independent, the expression for any arbitrary moment does not take up the simple form given above. One needs to account for the presence of higher order correlations amongst the random variables. This fact needs to be accounted while evaluating each of the moment.\\
\indent Specifically while calculating the moments of the BipoSH coefficients we find that the terms appearing in the linear combination have non-linear correlations. We evaluate the correction to the moments due to these non-linear correlations. We find that there is no correction to the variance as the terms involved in the linear combination turn out to be linearly uncorrelated. The corrected kurtosis is derived to have the following form,
\begin{eqnarray}
\bar \mu_4 = \tilde \mu_4 + \frac{3[\sum_{i}^{N}(K^{i}_{2})^{2}+2\sum_{i\ne j}E[X^{2}_{i}X^{2}_{j}]]}{(\sum_{i}^{N}K^{i}_{2})^{2}},
\end{eqnarray}
where second term is the correction term. In the above expression $K_{i}$ is the cumulant of the $i^{th}$ term and
$X_{i}$ and $X_{j}$ are the $i^{th}$ and $j^{th}$ terms in the summation.

The calculation for correction for higher order moments becomes very tedious, hence we restrict ourselves to calculating corrections for moments only upto kurtosis.
\end{widetext}

\end{document}